\documentclass[12pt]{article}
\usepackage{amsmath}
\usepackage{amssymb}
\usepackage{bbm}
 \usepackage[hmargin=3.54cm,tmargin=3.7cm,bmargin=3.7cm]{geometry}
 \usepackage{color}

\numberwithin{equation}{section}

 \usepackage{graphicx,psfrag}%

\usepackage{graphicx}

\usepackage{pict2e}%

 \usepackage{pdfsync}

\newcommand{\MAT}[1]{\left(\begin{array}{*#1c}}
\newcommand{\mat}{\end{array}\right)}
\newcommand{\qed}{\leavevmode\unskip\nobreak\penalty200\hskip2pt\null
\nobreak\hfill\rule{1.1ex}{1.1ex}%\parfillskip=0pt
\medbreak }

\newcommand{\N}{\mathbb{N}}
\newcommand{\Z}{\mathbb{Z}}
\newcommand{\R}{\mathbb{R}}
\newcommand{\Id}{\mathbbm{1}}

\newcommand{\DF}{\Longleftrightarrow}

\newcommand{\AR}{{\cal A}}

\newcommand{\LR}{{\cal L}}

\newcommand{\PR}{{\cal P}}

\newcommand{\WR}{{\cal W}}

\newcommand{\BX}{{\mathbb X}}
\newcommand{\BY}{{\mathbb Y}}

\newcommand{\iy}{\infty}
\newcommand{\pl}{\partial}

\newcommand{\bT}{{\mathbf T}}
\newcommand{\bt}{{\mathbf t}}

\newcommand{\bn}{{\mathbf n}}

\newcommand{\un}{\mbox{1\hspace{-3.2pt}I}}

\newenvironment{proof}{\medskip\noindent{\it Proof:\/} }{\qed}

\newcommand{\om}{\omega}
\newcommand{\vp}{\varphi}

\newcommand{\vr}{\varepsilon}

\newcommand{\BR}{{\mathbb R}}

\newcommand{\lb}{\lambda}

\newcommand{\dis}{\displaystyle}
\newcommand{\T}{{\mathrm T}}

\newcommand{\rr}{\operatorname{tr}}

\def\be#1\ee{\begin{equation}#1\end{equation}}
\def\bea#1\eea{\begin{eqnarray}#1\end{eqnarray}}
\def\bean#1\eean{\begin{eqnarray*}#1\end{eqnarray*}}

\newtheorem{definition}{Definition}[section]
\newtheorem{theorem}[definition]{Theorem}
\newtheorem{lemma}[definition]{Lemma}

\newtheorem{proposition}[definition]{Proposition}
\newtheorem{remark}[definition]{Remark}

\catcode `!=11

\newdimen\squaresize
\newdimen\thickness
\newdimen\Thickness
\newdimen\ll! \newdimen \uu! \newdimen\dd! \newdimen \rr! \newdimen
\temp!

\def\sq!#1#2#3#4#5{%
\ll!=#1 \uu!=#2 \dd!=#3 \rr!=#4
\setbox0=\hbox{%
 \temp!=\squaresize\advance\temp! by .5\uu!
 \rlap{\kern -.5\ll!
 \vbox{\hrule height \temp! width#1 depth .5\dd!}}%
 \temp!=\squaresize\advance\temp! by -.5\uu!
 \rlap{\raise\temp!
 \vbox{\hrule height #2 width \squaresize}}%
 \rlap{\raise -.5\dd!
 \vbox{\hrule height #3 width \squaresize}}%
 \temp!=\squaresize\advance\temp! by .5\uu!
 \rlap{\kern \squaresize \kern-.5\rr!
 \vbox{\hrule height \temp! width#4 depth .5\dd!}}%
 \rlap{\kern .5\squaresize\raise .5\squaresize
 \vbox to 0pt{\vss\hbox to 0pt{\hss $#5$\hss}\vss}}%
}%end of \hbox
 \ht0=0pt \dp0=0pt \box0
}%end of \sq!

\def\vsq!#1#2#3#4#5\endvsq!{\vbox to \squaresize{\hrule
width\squaresize height 0pt%
\vss\sq!{#1}{#2}{#3}{#4}{#5}}}

\newdimen \LL! \newdimen \UU! \newdimen \DD! \newdimen \RR!

\def\vvsq!{\futurelet\next\vvvsq!}
\def\vvvsq!{\relax
  \ifx     \next l\LL!=\Thickness \let\continue=\skipnexttoken!
  \else\ifx\next u\UU!=\Thickness \let\continue=\skipnexttoken!
  \else\ifx\next d\DD!=\Thickness \let\continue=\skipnexttoken!
  \else\ifx\next r\RR!=\Thickness \let\continue=\skipnexttoken!
  \else\def\continue{\vsq!\LL!\UU!\DD!\RR!}%
  \fi\fi\fi\fi
  \continue}

\def\skipnexttoken!#1{\vvsq!}

\def\place#1#2#3{\vbox to 0pt{\vss
\rlap{\kern#1\squaresize
  \raise#2\squaresize\hbox{$#3$}}
\vss}}

\catcode `!=12

\long\def\symbolfootnote[#1]#2{\begingroup%
\def\thefootnote{\fnsymbol{footnote}}\footnote[#1]{#2}\endgroup}

\begin{document}

\begin{titlepage}
\begin{flushright}
%CRM-???? (2009)\\
%nlin.SI/05xxxxxx
\end{flushright}
%\vspace{0.2cm}
\begin{center}
\begin{Large}
\textbf{Nonlinear PDEs for Fredholm determinants arising from string equations}
\end{Large}

\bigskip
M. Adler\symbolfootnote[2]{Department of Mathematics, Brandeis University, Waltham, Mass 02454, USA. adler@brandeis.edu. The support of a National Science Foundation grant \# DMS-07-00782 is gratefully acknowledged}\hspace{0.5cm}
M. Cafasso\symbolfootnote[3]{LUNAM Universit\'e, LAREMA, Universit\'e d'Angers,
2 Bd. Lavoisier 49045 Angers, France. cafasso@math.univ-angers.fr. The hospitality of the Max Planck Institute for Mathematics in Bonn is gratefully acknowledged.}\hspace{.5cm}
P. van Moerbeke\symbolfootnote[4]{D\'epartement	de	Math\'ematiques, Universit\'e Catholique	de Louvain, 1348 Louvain-la-Neuve, Belgium and Brandeis University, Waltham, Mass 02454, USA. pierre.vanmoerbeke@uclouvain.be. The support of a National Science Foundation grant \# DMS-07-00782, a European Science Foundation grant (MISGAM), a Marie Curie Grant (ENIGMA), Nato, FNRS and Francqui Foundation grants is gratefully acknowledged.}
\end{center}
\bigskip
%%%%%%%%%%%%%%%%  Abstract  %%%%%%%%%%%%%%%%
\begin{center}{\bf Abstract}
\end{center}
\qquad String equations related to 2D gravity seem to provide, quite naturally and systematically, integrable kernels, in the sense of Its-Izergin-Korepin and Slavnov. Some of these kernels (besides the ``classical'' examples of Airy and Pearcey) have already appeared in random matrix theory and they have a natural  Wronskian structure, given by one of the operators in the string relation $[L^\pm, Q^\pm]=\pm 1$, namely $L^\pm$. The kernels are intimately related to wave functions for Gel'fand-Dickey reductions of the KP hierarchy. The Fredholm determinants of these kernels also satisfy Virasoro constraints leading to PDEs for their log derivatives, and these PDEs depend explicitly on the solutions of Painlev\'e--like systems of ODEs equivalent to the relevant string relations. We give some examples coming from critical phenomena in random matrix theory (higher order Tracy--Widom distributions) and statistical mechanics (Ising models).
\end{titlepage}
\tableofcontents

\section{Introduction}

String equations have been introduced in the context of $(p,q)$ minimal models coupled to gravity by Douglas \cite{Douglas1} in 1990; their connections with different areas of the theory of integrable systems have been studied, after Douglas himself, by many different authors, see \cite{DFGZJ} and references therein. Kac and Schwarz, in particular, related string equations to the Sato's Grassmannian formulation of KP theory \cite{KacSchwarz,SchwarzSE}, while Moore provided analogous connections with the matrix  Lax formulation of Drin'feld--Sokolov hierarchies and, more importantly,  with the theory of isomonodromic deformations \cite{MooreGeometrySE, MooreMatrix}. More recently, some connections between the whole set of $(p,q)$ string equations and the scaling limit of Christoffell--Darboux kernels in multi--matrix models have been suggested in \cite{BergereEynard}.\footnote{Curiously enough, while the connection between the scaling limit of random matrices and string equations is known in the physics 
literature since the nineties, we are unable to make reference to any work giving, even conjecturally, the general form of the scaling limit of Christoffel--Darboux kernels in terms of the aforementioned string equations.}

The purpose  of this paper is to show how the basic KP integrable structure behind string relations leads to PDEs for the log of the Fredholm determinants of intrinsically associated kernels, the PDEs themselves containing explicitly the solutions to the ``Painlev\'e-like'' equations derived from the string relations.
To be precise, consider the string relations for the following differential operators of orders\footnote{In this paper the role of $p$ and $q$ have been interchanged with respect to the usual convention in physics literature (see for instance \cite{Douglas1}). We apologize for the inconvenience, but we preferred to remain consistent with the notation adopted in \cite{ACvM1}.} $p$ and $q$ (relatively prime) in $D:=\dis\frac{\partial}{\partial x}$:
\bea
	\left[\LR^\pm_{p;\bT_q},Q^\pm_{p;\bT_q}\right] &=& \pm 1,\label{introSE}\\
	\LR_{p;\bT_q}^\pm := D^p + \sum_{i = 0}^{p-2}\theta_i^\pm D^i,&\,& Q^\pm_{p;\bT_q} := \sum_{\ell = 1}^q T_{p+\ell}\left(\LR_{p;\bT_q}^{\pm\frac{\ell}{p}}\right)_+,
\eea
where $\bT_q := (T_{p+1},\ldots,T_{p+q})$ are some constants, $\LR_{p;\bT_q}^-$ is the formal adjoint of $\LR_{p;\bT_q}^+$ and  analogously for $Q^\pm_{p;\bT_q}$.\\

As it is known in the context of isomonodromic deformations \cite{MooreMatrix}, these string relations are solved in $\theta^+_0,\ldots,\theta_{p-2}^+$ by a ``Painlev\'e--like'' system of ODEs of the form\footnote{Of course analogue formulas can be written also for the variables $\theta^-_0,\ldots,\theta_{p-2}^-$.} 
\be
	\sum_{\ell = 1}^q T_{p+\ell}\left(\frac{\delta H_\ell^p}{\delta \theta_0^+},\ldots,\frac{\delta H_\ell^p}{\delta \theta_{p-2}^+}\right) + \bn = 0,
\label{introPSE}\ee
where 
\be
	H_\ell^p :=\frac{p}{p+\ell}\mathrm{tr}(\LR_{p;\bT_q}^{+\frac{p+\ell}p}),
\ee
``tr'' denotes the Adler's trace \cite{AdlerTrace}
$$\mathrm{tr}\left(\sum_i a_iD^i\right):=D^{-1}a_{-1},$$
and $\bn$ is the generic solution of the equation
$J_1\bn = \rm e_1^\T,$
where $J_1$ is the $(p-1)\times (p-1)$ matrix giving the first symplectic structure of the Gel'fand--Dickey hierarchy (see Section 2.1).\\

Now denote with $\Psi_{p;\bT_q}^\pm$ the KP wave functions associated to the differential operators $\LR^\pm_{p;\bT_q}$, satisfying the eigenvalues equations
\be
	\LR^\pm_{p;\bT_q}(x,\bt;z)\Psi_{p;\bT_q}^\pm(x,\bt;z) = z^p\Psi^\pm_{p;\bT_q}(x,\bt;z)
\ee
together with the following asymptotic conditions at infinity
\be
	\Psi_{p;\bT_q}^\pm(x,\bt;z) = {\rm e}^{\pm(xz + \sum_{i = 0}^\infty t_iz^i)}\left(1+\mathcal O\left(\frac{1}{z}\right)\right)
\ee 
and additional equations describing the evolution with respect to $\bt:=(t_1,t_2,t_3,\ldots)$ (see Section 2.1).
The wave functions above, together with the string relations (\ref{introSE}), lead quite naturally to a Wronskian kernel (see (\ref{IIKSkernels})) defined by the equation:
\bea
	&&DK_{x;\bt}^{p;\bT_q}(\lambda,\lambda') :=\label{introkernels}\\
	\nonumber\\
	 &&\dis\frac{\Phi^-_{p;\bT_q}(x,\bt;\lambda^{\frac{1}p})\left(\LR^+_{p;\bT_q}\Phi^+_{p;\bT_q}(x,\bt;\lambda^{'\frac{1}p})\right)-\Phi^+_{p;\bT_q}(x,\bt;\lambda^{'\frac{1}p})\left(\LR^-_{p;\bT_q}\Phi^-_{p;\bT_q}(x,\bt;\lambda^{\frac{1}p})\right)}{i(\lambda'-\lambda)}.\nonumber
\eea
The $\Phi_{p;\bT_q}^\pm$ are properly renormalized wave functions, namely
\be
	\Phi_{p;\bT_q}^\pm(x,\bt;z) := \frac{1}{\sqrt{\pm 2\pi p z^{p-1}}}{\rm e}^{\pm\sum_{\ell = 1}^q \frac{p}{p+\ell}T_{p+\ell}z^\ell}\Psi^\pm_{p;\bT_q}(x,\bt;z).
\ee
Using this integrable picture (namely combining the KP bilinear equations with relevant Virasoro constraints) we show, giving several physically--relevant examples, how to derive non--linear PDEs for  the log of the Fredholm determinants on $L^2(\R)$:
\be
	\det\left(\Id - 2\pi\mu \chi_E K_{x,\bt}^{(p,\bT_q)}(\lambda,\lambda')\right)_{\mathrm{most}\, t_i = 0},
\ee
where $E$ is a collection of intervals with endpoints $\left\{a_i\right\}_{i=1}^r$ and $\chi_E$ is its indicator function. These PDEs will depends on the parameters $x$, the non--zero $t_i$, the constants $\bT_q = (T_{p+1},\ldots T_{p+q})$, and also on the operators
\be
	\partial := \sum_{i = 1}^r \frac{\pl}{\pl a_i},\quad \varepsilon := \sum_{i=1}^r a_i\frac{\partial}{\partial a_i},
\ee
and the variables $\theta_0^+,\ldots,\theta_{p-2}^+$ satisfying the Painlev\'e--like ODEs (\ref{introPSE}).
Such Fredholm determinants arise naturally in random matrix theory. The simplest case is for $q=1$; it has already been considered (for $p$ arbitrary,) in \cite{ACvM1}. Indeed, for $(p,q) = (2,1)$ and $(p,q) = (3,1)$, the kernels (\ref{introkernels}) correspond to the Airy and the Pearcey kernel respectively. As we will show in this paper, the case $(p,q)=(2,3)$ has been considered in \cite{ClaeysVanlessen2} and the generalization for $p=2$ and $q$ arbitrary in \cite{CIK}. At the moment we are not able to give a ``physical'' meaning to all the kernels defined by (\ref{introkernels}) or in particular their Fredholm determinants, but it is just natural to relate them to the $(p,q)$ kernels described in the introduction of \cite{BergereEynard} and conjecturally associated to critical phenomena in multi--matrix models.\\

For example, consider with Claeys and Vanlessen \cite{ClaeysVanlessen2} the unitary random matrix model with probability measure
\be
	\frac{1}{Z_n}{\rm e}^{-n\mathrm{Tr} V(M)}dM
\ee
where $V$ is a polynomial such that, in the large $n$ limit, the density of state $\rho$ behaves at the endpoint of an interval $x_0$ like $\rho \sim c|x-x_0|^{5/2}.$ This is achieved by setting
\be
	V(z) = \frac{1}{20}z^4-\frac{4}{15}z^3+\frac{1}5 z^2 + \frac{8}5 z +\alpha z + \beta(z^3 - 6z)
\ee
and letting $n\rightarrow\infty,\, \alpha,\beta \rightarrow 0$ in such a way that that $n^{6/7}\alpha \rightarrow c_1 x,\, n^{4/7} \beta \rightarrow c_2 t$, for some constants $c_1,c_2$, with $x$ and $t$ being parameters. Then the usual 2--point correlation kernel $K_n(x,y)$ (Christoffel--Darboux kernel) has a universal limit, namely
\be
	\lim_{n\rightarrow\infty} \frac{1}{cn^\frac{2}7}K_n\left(x_0+\frac{u}{cn^{\frac{2}7}},x_0+\frac{v}{cn^{\frac{2}7}}\right) = K^{(1)}(u,v;x,t) = iK_{x;0}^{(2,\bT_5)}(u,v)
\ee
with $\bT_5=(-t/2,0,0,0,1/30)$ and $K_{x;0}^{(2,\bT_5)}(u,v)$ is given by (\ref{introkernels}) (see section 4.1). By Proposition \ref{ClaeysVanlessen} this leads to the following PDE for the log of the Fredholm determinant:
$$U(E,x,t) := \log\det ( \un -2\pi\mu K^{(2;\bT_5)}_{x,{0}} \raisebox{1mm}{$\chi$}{}_{E} ),$$
namely
\be
\left\lbrace 60 \pl\pl_xU\!\! +\! 30t \pl_x^2U\!\! -\!\! 6\pl_t^2U\!\! +\! \pl_t\pl_x^2U\!\! +\! 6\pl_x^2U\pl_x\pl_tU\!\! +\! 6y\pl_x\pl_tU,\pl_x^2U\right\rbrace_x\!\!+6\!\left(\pl_x^2U\right)^2\!\pl_ty =0;
\label{6.64bisintro}\ee
with $y=y(x,t)$ the solution to the Painlev\'e equation $PI^2$ (the second member of the Painlev\'e I hierarchy), namely\footnote{The case of \cite{ClaeysVanlessen2} would correspond to $c = 0$ and $E = [s,\infty)$.}
\be
	\frac{1}6 y^3 + \frac{1}{24}(\pl_x y)^2 + \frac{1}{12}y\pl_x^2y + \frac{1}{240}\pl_x^4y-ty + x + c= 0,\nonumber
\ee
\be \ee
\be
	y(x,t) = \mp \left(6|x|\right)^{\frac{1}{3}} \mp 6^{\frac{2}{3}}t|x|^{-\frac{1}{3}} + \mathcal{O}\left(|x|^{-1}\right),\quad x\rightarrow \pm\infty.\nonumber
\ee
More generally Claeys, Its and Krasovsky in \cite{CIK} have considered the case where $\rho \sim c|x-x_0|^{(4k+1)/2}$ at the endpoint. In this case one should set
\be
	V(z) = \tilde V(z) + \sum_{j = 0}^{2k-1} \alpha_j V_j(x),
\ee
with $n \rightarrow \infty,\, \alpha_j \rightarrow 0$ appropriately. Then the 2-point correlation  kernel $K_n$ is  conjectured to have a universal limit
\be
	\lim_{n\rightarrow\infty} \frac{1}{cn^{\frac{2}{4k+3}}}K_n(x_0+\frac{u}{c n^{\frac{2}{4k+3}}}, x_0+\frac{v}{c n^{\frac{2}{4k+3}}}):= K^{(k)}(u,v;x,\bT_{4k+1}) = iK_{x,0}^{(2,\bT_{4k+1})},
\ee
with $\bT_{4k+1} = (T_3,0,T_5,0,T_7,\ldots,T_{4k+3})$.
The methods of section 5 would lead to a PDE for $\det(\Id-\chi_{[s,\infty)}K^{(k)}(u,v;x,\bT_{4k+1}))$, depending on the parameters $x ,s$, some of the $T_{2j+1}$ and the variable $\theta_0(x,\bT_{4k+1})$ satisfying the equation (belonging to the Painlev\'e I hierarchy)
\be
	2\sum_{j=1}^{2k+1} T_{2j+1}\omega_j\left(\frac{\theta_0}2\right) + x = 0
\ee
where $\omega_j$ are the Gel'fand--Dickey polynomials defined in (\ref{defGDpoly}). Very interestingly the results obtained in \cite{CIK} have been rederived recently \cite{AtAk} using the Lax operators related to orthogonal polynomials and their asymptotics in the double scaling limit.\\

The following two examples come from matrix models and statistical mechanics; both of them are taken from \cite{DFGZJ} and they relate Ising models to some multi--matrix models. First consider the so--called \emph{critical Ising model}, which has a realization as a two-matrix model possessing the string relations
                                                    $$[\LR^+_{3;\bT_4},Q^+_{3;\bT_4}] = 1,$$
with $\bT_4=(0,T_5,0,1)$
and with
\bea
	\LR_{3;\bT_4}^+ &=& \left((D^2 - u)^{\frac{3}{2}}\right)_+ + \frac{3}{2} w = D^3 - \frac{3}2 uD + \frac{3}4 (2w-u'),\nonumber\\
	Q^+_{3;\bT_4} &=&\!\!\!\! (\LR_{3;\bT_4}^{+\frac{4}3})_+ + T_5(\LR_{3;\bT_4}^{+\frac{2}3})_+.
\eea
Then $u$ and $w$ are a solution of the Painlev\'e-like system of equations
\be\left\{\begin{aligned}
	&\dis\frac{1}2 w''-\dis\frac{3}2 uw + \dis\frac{3}2 T_5 w + t_2 = 0,\\
	\\
	&\dis\frac{1}{12} u^{(iv)}-\dis\frac{3}4 uu''-\frac{3}{16} (u')^2 + \dis\frac{1}4 u^3 -\dis\frac{1}{4} T_5(3u^2 - u'') + \dis\frac{3}2 w^2 +x = 0,\end{aligned}\right.
\label{string43intro}\ee
and it is a consequence of Proposition \ref{criticalIsing} that the log of the Fredholm determinant
$$\left.V(E,x,t_2) := \log\det ( \un -2\pi\mu K^{(3;\bT_4)}_{x,{\bt}} \raisebox{1mm}{$\chi$}{}_{E} )\right | _{\underset{i\neq 2}{t_i = 0}}$$
satisfies the following PDE involving $u$ and $w$ (set $\partial_i := \frac{\partial}{\partial t_i}$):
\be
3T_5\pl_2\pl_x V -3\pl\pl_xV+\pl^3_x\pl_2V+6(\pl^2_xV)(\pl_x\pl_2V)-12u(\pl_x\pl_2V)+6w\pl^2_xV=0.
\label{6.64intro}\ee

The last case we consider is related to the so-called \emph{tricritical Ising model} (see again \cite{DFGZJ} and references therein) and is expressed via the string equation
$$[\LR^+_{4;\bT_5},Q^+_{4;\bT_5}] = 1,$$
with $\bT_5=(0,0,0,0,1)$
and with
\bea
	\LR_{4;\bT_5}^+& = &(D^2 - u)^2 + wD + Dw +v,\nonumber\\
	\label{LaxtriIsingintro}\\
	Q^+_{4;\bT_5}& =&(\LR_{4;\bT_5}^{+\frac{5}4})_+, \nonumber
\eea
with $u,v$ and $w$ satisfying the Painlev\'e-like system of equations (\ref{stringtriIsing}). It is a consequence of Proposition 5.5 that the log of the Fredholm determinant
$$\left.W(E,x,t_2,t_3) := \log\det ( \un -2\pi\mu K^{(4;\bT_5)}_{x,{\bt}} \raisebox{1mm}{$\chi$}{}_{E} )\right | _{\underset{i\neq 2,3}{t_i = 0}}$$
satisfies the following PDE depending on $u,v$:
\bea
 \frac{1}5 \pl_x^6W-4\pl_3^2W+2\pl_x^2\pl_3W+12\left(\pl_x^2W\right)^3+6\left(\pl_x^4W\right)\left(\pl_x^2W\right)+12\left(\pl_x^2W\right)\left(\pl_x\pl_3W\right)\nonumber \\
-72 u \left(\pl_x^2W\right)^2-12 u (\pl_x^4W)-24 u (\pl_x\pl_3W)+\left(9v+72u^2-18u''\right) \pl_x^2W=\frac{36}5\pl\pl_x W. \nonumber\\
\label{PDEtriIsingintro}\eea
The PDEs (\ref{6.64bisintro}),(\ref{6.64intro}) and (\ref{PDEtriIsingintro}) are new. The technique used to derive the PDEs, given in section 3, is a generalization of the methods of \cite{ASvM,ACvM1}. Section 2 contains a quick review of KP theory from \cite{DJKM1,DJKM2} and \cite{ASvM,ASvM2,ACvM1}, plus a quick discussion with proofs of needed facts from the theory of string equations, since it is not easy to find actual proofs in the literature (see also \cite{MooreGeometrySE}). In section 3 we introduce the $(p,q)$ kernels, put them into a useful Wronskian form (showing \emph{en passant} that they are integrable kernels in the sense of \cite{IIKS}) and, moreover, derive the Virasoro relations we shall need to derive the PDEs. Section 4 provides the examples we discuss in the paper and finally in section 5 we derive the PDEs for the log of the Fredholm determinants involving the kernels coming from section 3 restricted to our examples of section 4. In this paper, as opposed to \cite{ACvM1}, we choose not 
to state a general PDE theorem, but rather explain a general method and implement it in a few well chosen examples. Given the diversity of string relations, that seemed the most transparent way to proceed.\\
              
There are certainly many open questions remaining, here we present a list of the most important ones.
\begin{itemize}
 	\item It would be nice to use the PDEs to derive useful asymptotic information about the Fredholm determinants, but that would probably require some insight into the solutions of our Painlev\'e-like equations.
	\item The theory of Eynard--Orantin symplectic invariants have been applied to the study of integrable kernels in many different articles \cite{BergereEynard,determinantalformulae,CM,GBE}. Nevertheless, till now, just the case of hyperelliptic curves (and related $(2\times2)$ Lax systems) have been studied. An interpretation of our kernels in terms of symplectic invariants could give some insight on the asymptotics of the Fredholm determinants (see previous point), as it has been done for the Tracy--Widom distribution in \cite{GBE}.
	\item In \cite{CIK} the authors proved that the Fredholm determinants related (in our language) to $(2,q)$ string equations are all expressible in terms of the so--called Hasting--Mc Leod solution of PII, already appearing in the expression of the Tracy--Widom distribution. In would be extremely interesting to re--derive their results with the formalism presented here. After the first version of this article appeared, the results obtained in \cite{CIK} have been rederived in \cite{AtAk}  using the Lax formalism. As noticed by Akemann and Atkin, their work give a partial answer to this question that deserve further investigations.
	\item The physical significance of the Fredholm determinants going with the Ising models is, to our knowledge, not known. More generally it would be interesting to find the physical significance of the whole class of kernels we introduced. This appear to be an ambitious project, since even for the $(p,1)$ cases presented in \cite{ACvM1}, this result has not been achieved.
	\item We do not know (as it is customary when differential equations are obtained with the method developed in \cite{ASvM,ACvM1}), if our PDEs posses a (properly formulated) Painlev\'e property. Also a Lax formulation of these PDEs is missing.    
\end{itemize}
\section{KP theory and $(p,q)$--string equations}

The basic tools we will need from the theory of integrable systems is the Sato's Grassmannian description of the KP hierarchy, Gel'fand--Dickey reductions (see for example \cite{DJKM1,DJKM2,ASvM, ASvM2, solitons}) and their relations with the solutions of the so--called $(p,q)$ string equations, i.e. equations of type
\be
	[\mathcal D_p,\mathcal D_q] = 1,
\label{SE}\ee
where $\mathcal D_p$ and $\mathcal D_q$ are differential operators of order $p$ and $q$ (see for instance \cite{SchwarzSE,MooreMatrix,MooreGeometrySE} and references therein).
In this section we briefly recall the notations and some results we used in \cite{ACvM1} and add (in the next two subsections) some facts related to the equation (\ref{SE}).\\

The KP hierarchy is a (infinite) set of integrable PDEs for a function $\tau(\bt)$ depending on a (infinite) set of variables $\bt:= (t_1,t_2,t_3,\ldots)$.
The whole set of equations is encoded in the famous bilinear identity
\be
	\oint_{\infty} \tau({\bf t}-[z^{-1}])\tau({\bf t'}+[z^{-1}]){\rm e}^{\sum_i^{\infty}(t_i-t'_i)z^i} = 0.
\label{BHE}\ee
In the expression above we denoted $[z] := (z,z^2/2,z^3/3,\ldots)$ (and similarly for $z^{-1}$); the  equations of the hierarchy are obtained expanding the integrand as a formal Laurent series about $z^{-1} = 0$ and then taking the (formal) residue about $z = \infty$.   

The hierarchy can also be written in Lax form as follows. Let us start setting
\be
	\overline{\mathbf t} := {\mathbf t} + x\mathrm{e}_1 = (x+t_1,t_2,t_3,\ldots),\quad 
	\partial_{\mathbf t} :=\left(\frac{\partial}{\partial t_1}, \frac{1}{2}\frac{\partial}{\partial t_2}, \frac{1}{3}\frac{\partial}{\partial t_3},\ldots\right)
\ee  
and denoting with  $p_i({\mathbf t})$ the classical Schur polynomials defined by:

\be\label{Schur}
	{\rm e}^{\sum_{i=1}^\infty t_iz^i}  = \sum_{i=0}^\infty z^i p_i({\mathbf t});
\ee
we also define  the wave operator $W=W(\bar{\mathbf t})$
\be
	W:=\frac{\tau(\bar{\mathbf t}-[D^{-1}])}{\tau(\bar{\mathbf t})}{\rm e}^{\sum_1^{\iy}t_iD^i}= 
	\left(\sum^{\iy}_{j=0}\frac{p_j(-\pl_{\mathbf t})\tau(\bar{\mathbf t})}{\tau(\bar{\mathbf t})}D^{- j}\right){\rm e}^{\sum_1^{\iy}t_iD^i}.
\label{waveoperator}\ee
Denoting with $H^*$ the formal adjoint of a given pseudo--differential operator $H$ (the formal adjoint acts through the formula $(a(x)D^j)^*:= (-D)^ja(x),\; j\in\Z$ and linearity) we can define the wave function $\Psi^+$ and Lax operators $L^+,M^+$ together with their adjoints (denoted with the minus sign)

\be
\begin{array}{lll}
L^+:=WDW^{-1},&M^+:=WxW^{-1},&\Psi^+: =We^{xz}\\
\\
L^-:=(W^{-1})^*(-D)W^*,&M^-:=(W^{-1})^*xW^*,&\Psi^-:=(W^{-1})^*e^{- xz}.
\end{array}
\label{7}\ee
One should think of $L^\pm$ and $M^\pm$ as the ``dressing'' of $\pm D$ and $x$ while the wave functions $\Psi^\pm$ are ``dressing'' of ${\rm e}^{\pm xz}$; this leading to the following relations, which result from ``undressing'' the operators and functions, to wit:

\be
\begin{array}{lll}
L^\pm\Psi^\pm=z\Psi^\pm,&M^\pm\Psi^\pm =\pm\dis\frac{\pl}{\pl z}\Psi^\pm,&[L^\pm,M^\pm]=\pm1.
\end{array}
\label{8}\ee
%\textcolor{red}{Also the definition of the wave operator $W$ (eq. (\ref{waveoperator})) together with (\ref{7}) gives back the celebrated Sato's formula expressing the wave functions in terms of the tau function, namely
%\be
%	\Psi^\pm(\bar{\mathbf t};z) := {\rm e}^{\pm xz \pm \sum_{i=1}^\infty t_iz^i}\dfrac{\tau\left(\bar{\mathbf t}\mp [z^{-1}]\right)}{\tau(\bar{\mathbf t})}.
%\label{SatoFormula}\ee}
Equations (\ref{7}), together with the bilinear identity (\ref{BHE}), gives Sato's formula for the wave function and its deformation equations\footnote{Here and below $L^{-i}$ means $(L^-)^i$ and not $L$ to the power $-i$ (on the other hand we did not introduce any operator $L$), while, given a pseudo--differential operator $A:=\sum_i a_iD^i$, we denote $A_+ = \sum_{i\geq 0} a_iD^i,\;A^- = A-A^+$.}
\be
	\Psi^\pm(\bar{\mathbf t};z) = {\rm e}^{\pm \left(xz + \sum_{i = 1}^\infty t_iz^i\right)}\frac{\tau(\bar{\mathbf t}\mp [z^{-1}])}{\tau(\bar{\mathbf t})},\quad\quad \frac{\pl}{\pl t_i}\Psi^\pm =\pm(L^{\pm i})_+\Psi^\pm, \quad i\in\Z_+,\label{9'}
\ee
 and these deformations, finally, give as compatibility conditions the Lax equations:
\be
 \frac{\pl}{\pl t_i}L^\pm=\left[\pm(L^{\pm i})_ +,L^\pm\right] \quad i \in \Z_+,
\label{9}\ee
leading to the famous KP equation describing shallow water waves in $\R^2$ for $q = \frac{\pl^2}{\pl x^2}\log\tau(\bar{\bt})$, with $(t_2,t_3) = (y,t)$:
$$q_{xxxx}+12q_x^2+12qq_{xx}+3q_{yy}-4q_{xt} = 0.$$

%Sato observed that the KP flows linearize on an infinite--dimensional Grassmannian $\mathrm{Gr}$ whose elements are linear spaces $\WR$ of formal Laurent series, i.e. typically we have\footnote{As the expert readers will have noticed, we are just describing the so--called ``big cell'' of the Sato's Grassmannian, which is enough for our purposes.}
%\be
%	\WR = \mathrm{span}\left\{\varphi_n(z) = z^n + \sum_{-\infty<i\leq n-1} a_{ni} z^i,\;a_{ni}\in\BC\right\}_{n=0}^\infty .
%\label{basis}\ee
Following Sato we associate, to any KP solution, two subspaces $\WR^\pm$ in Sato's Grassmannian of vector spaces spanned by a formal basis of the form $$\left\{z^{s_i}\left(\sum_{j\leq 0}a_{ij}z^j\right)\right\}_{i \geq 0},\quad s_i = i\;\mathrm{eventually}$$  through the formula
\be
\begin{array}{lll}
\WR^\pm&:=& {\rm span}_{i\geq  
0}   \{D^i\Psi^\pm(x,0;z)\};
\end{array}
\label{11}\ee
these linear spaces $\WR^\pm$ are ${\mathbf t}$--deformed by the KP flows via
\be
	\WR^\pm({\mathbf t}) = {\rm e}^{\mp\sum_{i=1}^\infty t_iz^i}\WR^\pm.
\label{GrassmannFlow}\ee
We also need, in the sequel, the following mapping from $z$--operators $A$ to $x$--operators $\mathcal P_A$ given by  

\be
\begin{array}{lll}
A^+\Psi^+&:=&\dis\sum_{-\iy <i<\iy}\sum_{j\geq 0}c_{ij}z^i\left(\dis\frac{\pl}{\pl  
z}\right)^j\Psi^+=\sum_{i,j}c_{ij}(M^{+})^j(L^{+})^i\Psi^+ =:\PR_{A^+}^+\Psi^+,\\
\\
A^-\Psi^-&:=&\dis\sum_{-\iy <i<\iy}\sum_{j\geq 0}c_{ij}\left(-\dis\frac{\pl} 
{\pl z}\right)^j z^i\Psi^- =\sum_{i,j}c_{ij}(L^{-})^i(M^{-})^j\Psi^-=:\PR_{A^-}^- 
\Psi^-,
\end{array}
\label{10}\ee
where the equalities are proven using (\ref{8}). The following implication follows from (\ref{10}) and relates the invariance properties of $\WR^\pm$ with regard to differential operators in $z$ to properties of the associated differential operator in $x$.
\be
A^+\WR^+\subset \WR^+\DF \PR_{A^+}^+=(\PR_{A^+}^+)_+ \DF A^-\WR^-\subset \WR^-.
\label{12}\ee

\subsection{p--reduced KP--hierarchies and $(p,q)$-string equations.}

The KP hierarchy contains, as reductions, the so--called Gel'fand--Dickey hierarchies, which we shall refer to as the p--reduced KP hierarchies. The following  lemma is well known and it has been already proven, with this notation, in \cite{ACvM1}:
\begin{lemma}\label{equiv}
	Let $p$ be a non--negative integer. Given a point $\WR^+$ in the Sato's Grassmanian and let $L^+$ and $\tau$ be the Lax operator and the tau function of the corresponding solution of the KP hierarchy. The following conditions are equivalent and are conserved along the KP flows.
	\begin{enumerate}
		\item $z^p\WR^+\subseteq \WR^+$
		\item $L^{+p} = (L^{+p})_+$
		\item $\tau$ does not depend on $t_{np}$ for any $n \geq 1$, modulo a removable factor of the form ${\rm e}^{\sum_{i=1}^\infty c_nt_{np}}$.
	\end{enumerate}
\end{lemma}

This lemma leads us to the following definition:

\begin{definition}
The p--reduced KP--hierarchy is the KP--hierarchy supplemented with one of the (equivalent) conditions 1. 2. or 3. of Lemma \ref{equiv}. 
\end{definition}
In the case of p--reduced KP hierarchies the relevant Lax operator, rather then being $L^+$, is its $p^{th}$ power $L^{+p}$, since the latter is a differential operator. Indeed, upon using the Lax formulation of the KP hierarchy and setting

$$\LR^\pm := L^{\pm p},$$
one finds equations
\be
	\LR^\pm \Psi^\pm = z^p \Psi^\pm,\quad \frac{\pl \Psi^{\pm}}{\pl t_i}=(\pm\LR^{\pm{i}/{p}})_+\Psi^\pm,\quad i\in\Z_+
\label{pKP0}\ee
and their compatibility conditions give the Lax formulation of the p--reduced KP hierarchy:
\be
	\frac{\pl\LR^{\pm}}{\pl  t_i}=\bigl [(\pm\LR^{\pm{i}/{p}})_+,\LR \bigr],\quad i\in\Z_+.
\label{pKP}\ee

In \cite{ACvM1} we used, in order to develop our theory, some particular solutions of p--reduced KP hierarchies satisfying an additional invariance property. Let's start defining a $z$--operator 
\be
	\AR_p^\pm(z) := z\pm\dis\frac{1}{pz^p}\left(z\dis\frac{\pl}{\pl z}-\dis\frac{p-1} {2}\right);
\label{duesedici}\ee
it is easy to check that $\AR_p^\pm$ satisfies the following condition:
$$[\AR_p^\pm(z), z^p] = \pm 1.$$
The operator $\AR_p$ is sometimes called the Kac--Schwarz operator after the seminal paper \cite{KacSchwarz}. As a matter of fact it determines uniquely, by invariance, a unique point in the Sato's Grassmanian corresponding to a solution of the p--reduced KP hierarchy, as specified in the Theorem below. Parts of this theorem already appeared in \cite{KacSchwarz} and \cite{ASvM}, but perhaps with sketchy proofs for the case $p > 2$; in \cite{ACvM1} a complete proof can be found.
\begin{theorem} \label{theo1}%$W\in Gr^{(0)}$ is uniquely\footnote{$W\in Gr^{(0)}$  
%if and only if $W={\rm span}_{i\geq 0}\{z^i(1+{\bf O}(\frac{1}{z}))\} 
%$.} determined by 
%
The invariance conditions
\be\begin{aligned}
& z^p\WR^+ \subset \WR^+,\quad \AR^+_p\WR^+\subset \WR^+,\\
%& z^pW^*\subset W^*,\quad\AR_p^* W^*\subset W^*
 \end{aligned} \label{16}\ee
determine uniquely a plane $\WR^+_p\in \mathrm{Gr}$, which moreover uniquely determines $\WR^-_p\in\mathrm{Gr}$ by the relations
 \be
z^p\WR^-\subset \WR^-,\quad\AR_p^- \WR^-\subset \WR^- . \label{17} \ee
The $\WR^\pm_p$ are linearly generated by the eigenfunctions $\vp_p^\pm$ of the operators $\left(\AR_p^{\pm}\right)^p$, namely:
\be
\WR^\pm_p={\rm span}_{i\geq 0}\{(\AR_p^\pm)^i\vp_p^\pm\}
\label{18}\ee
with
 \be (\AR_p^{\pm})^p\vp_p^\pm=z^p\vp_p^\pm,\quad\vp_p^\pm(z)=1+\sum_1^{\iy}\frac{a_i^\pm}{z^i},  \label{22}\ee
the latter which uniquely determines $\varphi_p^{\pm}(z)$.\\
The corresponding p--reduced KP wave functions $\Psi^\pm_p(x,0;z)$ are then uniquely specified by
\be
\begin{array}{ll}
\Psi^\pm_p(0,0;z)=\vp_p^\pm(z)\\
\\
\AR_p^\pm(z)\Psi^\pm_p(x,0;z)=\pm\dis\frac{\pl}{\pl x}\Psi^\pm_p(x,0;z).
\end{array}
  \label{19}\ee
\end{theorem}
Given our subspaces $\WR_p$ we want to move them along the KP flows so to get solutions of p--reduced KP equations nicely related to the string equation (\ref{SE})\footnote{This process, at the level of tau functions, corresponds to merely shifting $\bt$, hence the choice of an origin in $\bt$ is relevant to this paper.}
\begin{definition}\label{shifteddefinition}
	Let $q\in\N$ and $\bT_q := (T_{p+1},\ldots,T_{p+q})\in\R^{q}$ a vector such that $T_{np} = 0,\,\forall\, n >0$ and $T_{p+q} \neq 0$.   Given $\WR_p^\pm$ uniquely determined by the invariance conditions (\ref{16}),(\ref{17}) we denote the KP--time deformed subspaces $\WR_{p;\bT_q}^\pm$ by
	\be
		\WR_{p;\bT_q}^\pm := {\rm e}^{\pm \frac{p}{p+1}z^{p+1}\mp\sum_{\ell =1}^q \frac{p}{p+\ell} T_{p+\ell} z^{p+\ell} }\WR_{p}^\pm.
	\label{shiftedsubspaces}\ee
\end{definition}
Note that, in this setting, $\WR_p^\pm = \WR_{p;(1)}^\pm$. In the notation of (\ref{GrassmannFlow}) 
$$\WR_{p;\bT_q}^\pm = \WR_p^\pm(\bt),\;\mathrm{with}\; t_{p+\ell} = \frac{p}{p+\ell}(T_{p+\ell}-\delta_{\ell,1}),\;1 \leq \ell \leq q,\quad t_i = 0\;\mathrm{otherwise.} $$ 
In the following we denote with $\LR_{p;\bT_q}^\pm$ the Lax operators, as in (\ref{pKP0}), corresponding to these subspaces through the Sato's theory. The following theorem is a generalization of a result in \cite{ACvM1}, which itself was based on the constructions in \cite{KacSchwarz}. For a further study of the manifold of string relations, one should consult the paper \cite{SchwarzSE}.
\begin{theorem}
	The subspaces $\WR_{p;\bT_q}^\pm$ satisfy the invariance conditions
	\be
		z^p \WR^\pm_{p;\bT_q} \subset \WR_{p;\bT_q}^\pm,\quad \mathcal A_{p;\bT_q}^\pm\WR_{p;\bT_q}^\pm \subset \WR_{p;\bT_q}^\pm
	\label{generalinvariance}\ee
	with 
	$$ \mathcal A_{p;\bT_q}^\pm = \sum_{\ell = 1}^q T_{p+\ell}z^\ell\pm \frac{1}{pz^p}\left(z\frac{\pl}{\pl z}-\frac{p-1}{2}\right).$$ 
	The corresponding Lax operators $\LR^\pm_{p;\bT_q}$ satisfies the string equations
	\be
		[\LR_{p;\bT_q}^\pm,Q_{p;\bT_q}^\pm] = \pm 1,\,{\mathrm{with}}\; Q^\pm_{p;\bT_q}:=\sum_{\ell = 1}^q T_{p+\ell}(\LR_{p;\bT_q}^{\pm\ell/p})_+.
	\label{SE2}\ee
\end{theorem}
Here again observe that, in analogy with the notation adopted for the subspaces $\WR^\pm_{p;\bT_q}$, we have $\mathcal A_p^\pm = \mathcal A_{p;(1)}^\pm$.\\

\proof
In (\ref{generalinvariance}) just the second equation needs to be proven. We observe that, using straightforward computations, the following commutation relations are easily verified:
\be
	\mathcal A_{p;\bT_q}^\pm{\rm e}^{\pm \frac{p}{p+1}z^{p+1}\mp\sum_{\ell =1}^q \frac{p}{p+\ell} T_{p+\ell} z^{p+\ell} } =  {\rm e}^{\pm \frac{p}{p+1}z^{p+1}\mp\sum_{\ell =1}^q \frac{p}{p+\ell} T_{p+\ell} z^{p+\ell} }\mathcal A_{p;(1)}^\pm.
\label{commutation}
\ee
On the other hand, from the previous theorem, we already have
\be
	\mathcal A_{p;(1)}^\pm\WR_{p;(1)}^\pm \subset \WR_{p;(1)}^\pm;
\label{firstinvariance}
\ee
so that using (\ref{commutation})-(\ref{firstinvariance}) together with the definition (\ref{shiftedsubspaces}) of $\WR^\pm_{p;\bT_q}$ we get
\bea
	\mathcal A_{p;\bT_q}^\pm\WR_{p;\bT_q}^\pm &=& \mathcal A_{p;\bT_q}^\pm {\rm e}^{\pm \frac{p}{p+1}z^{p+1}\mp\sum_{\ell =1}^q \frac{p}{p+\ell} T_{p+\ell} z^{p+\ell} }\WR_{p;(1)}^\pm = \nonumber\\
	={\rm e}^{\pm \frac{p}{p+1}z^{p+1}\mp\sum_{\ell =1}^q \frac{p}{p+\ell} T_{p+\ell} z^{p+\ell} }\mathcal A_{p;(1)}^\pm\WR_{p;(1)}^\pm&\subset& {\rm e}^{\pm \frac{p}{p+1}z^{p+1}\mp\sum_{\ell =1}^q \frac{p}{p+\ell} T_{p+\ell} z^{p+\ell} }\WR_{p;(1)}^\pm = \WR_{p;\bT_q}^\pm. \nonumber
\eea
Now we have to prove the string equation. 
%\bean
%	\left[\LR^+_{p;\bT_q},\frac{M^+_{p;\bT_q}\LR_{p;\bT_q}^{+\frac{1-p}{p}}}{p}\right] 
%	=\left[\LR^+_{p;\bT_q},\frac{M^+_{p;\bT_q}\LR_{p;\bT_q}^{+\frac{1-p}{p}}}{p}-\frac{(p-1)}{2p}\left(\LR_{p;\bT_q}^+\right)^{-1}+\sum^q_{\ell = 1}T_{p+\ell}\LR_{p;\bT_q}^{+\frac{\ell}{p}}
 %\right]  = 1\nonumber
 %\eean
Using (\ref{generalinvariance}) combined with (\ref{12}), we get 
 $\PR^+_{\AR_{p;\bT_q}^+} = \left(\PR^+_{\AR_{p;\bT_q}^+}\right)_+$, hence\footnote{Here $M^\pm_{p;\bT_q}$ denote the Lax operators associated to $\WR^\pm_{p;\bT_q}$ via equations (\ref{7}).}:
 \bean
\pm 1&=& \left[\AR_{p;\bT_q}^\pm,z^p\right] =  \left[\PR_{z^p}^\pm,\PR^\pm_{\AR_{p;\bT_q}^\pm}\right] = \left[\PR_{z^p}^\pm,\left(\PR^\pm_{\AR_{p;\bT_q}^\pm}\right)_+\right] \\
 \\
 &=&\left[ \LR^\pm_{p;\bT_q},\left(\frac{M^\pm_{p;\bT_q}\LR_{p;\bT_q}^{\pm\frac{1-p}{p}}}{p}-\frac{(p-1)}{2p}\left(\LR_{p;\bT_q}^\pm\right)^{-1}+\sum^q_{\ell = 1}T_{p+\ell}\LR_{p;\bT_q}^{\pm\frac{\ell}{p}}\right)_+
 \right]\\
 \\
 &=&\left[\LR^\pm_{p;\bT_q},\sum^q_{\ell=1}T_{p+\ell}(\LR_{p;\bT_q}^{\pm\frac{\ell}{p}})_+\right]=\left[\LR^\pm_{p;\bT_q},Q^\pm_{p;\bT_q}\right],
 \eean
 where in the first and second line we have used (\ref{10}) and (\ref{12}). \qed

The string equations (\ref{SE2}) can be written as compatibility conditions for a Lax system. Let us define the normalized wave functions and wave vectors:
\bea
	\Phi_{p;\bT_q}^\pm(x,\bt;z) &:=& \frac{1}{\sqrt{\pm 2\pi p z^{p-1}}} {\rm e}^{\pm \sum_{\ell = 1}^q \frac{p}{p+\ell}T_{p+\ell}z^{p+\ell}}\Psi_{p;\bT_q}^\pm(x,\bt;z);\label{normalizedwf}\\
	\hat\Phi_{p;\bT_q}^\pm(x,\bt;z) &:=& \left(\Phi_{p;\bT_q}^\pm, D\Phi_{p;\bT_q}^\pm,\ldots, D^{p-1}\Phi_{p;\bT_q}^\pm\right)^\T.\label{normalizedvwf}
\eea
where $\Psi_{p;\bT_q}^\pm$ are the wave function canonically associated, through Sato's theory, to the subspaces $\WR_{p;\bT_q}^\pm$ (so that in particular $\Psi_{p;(1)}^\pm = \Psi_{p}^\pm$). Let us also set $$\LR_{p;\bT_q}^\pm = D^p + \sum_{i = 0}^{p-2}\theta_i^{\pm}D^i.$$
\begin{theorem}\label{ccSE}
	The normalized wave functions $\Phi_{p;\bT_q}^\pm$, satisfy the relations
	\be
		\LR_{p;\bT_q}^\pm \Phi_{p;\bT_q}^\pm = z^p \Phi_{p;\bT_q}^\pm, \quad Q_{p;\bT_q}^\pm \Phi_{p;\bT_q}^\pm = \pm \frac{\pl}{\pl (z^p)}\Phi_{p;\bT_q}^\pm ,
	\label{LaxNormalized}\ee
whose compatibility conditions are the string relations (\ref{SE2})
\be
	\left[\LR_{p;\bT_q}^\pm, Q_{p;\bT_q}^\pm \right] = \pm 1.
\ee
The relations (\ref{LaxNormalized}) are, in the usual fashion, equivalent to the first order system:
\be
	D\hat\Phi_{p;\bT_q}^{\pm} = U^\pm \hat\Phi^\pm_{p;\bT_q},\quad \pm\frac{\pl }{\pl (z^p)}\hat\Phi_{p;\bT_q}^{\pm} = V_{p,\bT_q}^\pm\hat\Phi_{p;\bT_q}^{\pm}.
\label{mpKP0}\ee
Here
\be
U^\pm:=\left[\begin{array}{ccccc}
			0 & 1 &  &  & {\bf O }\\
			 & \ddots & \ddots & &\\
			 & & \ddots & \ddots &\\
			 \bf{O} & & & \ddots & 1\\
			 z^p-\theta^\pm_0 & -\theta^\pm_1 & \ldots & -\theta^\pm_{p-2} & 0
		\end{array}\right]
\label{U}\ee
and $V_{p,\bT_q}^\pm$ are some $p\times p$ matrices whose coefficients are polynomials in the variables $\left\{z^p, D^k \theta_\ell^\pm; \, k \geq 0, \ell = 0,\ldots,p-2\right\}$, of degree at most $\left[\frac{q-1}{p}\right] +1$ in $z^p$, completely determined by $Q^\pm_{p;\bT_q}$. Consequently their compatibility relations are written in the form
\be
	\pm \frac{\pl}{\pl (z^p)}U^\pm - D V^\pm_{p;\bT_q} = \left[V^\pm_{p;\bT_q},U^\pm\right].
\label{matrixSE}\ee
Relations (\ref{LaxNormalized}) or (\ref{matrixSE}), along with the asymptotic relation as $z\rightarrow\infty :$
\be
	\Phi_{p;\bT_q}^\pm(x,0;z) = \frac{1}{\sqrt{\pm 2\pi p z^{p-1}}}{\rm e}^{\pm\left(xz + \sum_{\ell = 1}^q \frac{p}{p+\ell}T_{p+\ell}z^{p+\ell}\right)}\left(1+\mathcal O\left(\frac{1}z\right) \right),
\label{asymPhi}\ee
characterizes $\Phi^\pm_{p;\bT_q}(x,0;z)$ as an asymptotic series and through (\ref{normalizedwf}) uniquely determine $\WR_{p;\bT_q}^\pm$ of (\ref{shiftedsubspaces}), and hence $\LR_{p;\bT_q}^\pm$ and $Q_{p;\bT_q}^\pm$. 
\end{theorem}
\proof
Let us temporarily denote
$$C^\pm(z) := \frac{1}{\sqrt{\pm 2\pi p z^{p-1}}}{\rm e}^{\pm \sum_{\ell = 1}^q \frac{p}{p+\ell}T_{p+\ell}z^{p+\ell}}.$$
The first equation in (\ref{LaxNormalized}) comes from the equality
$$\LR_{p;\bT_q}^\pm \Psi_{p;\bT_q}^\pm = z^p \Psi_{p;\bT_q}^\pm;$$ 
observing that $\LR_{p;\bT_q}^\pm$ acts trivially on $C^\pm(z)$.
The second equation comes from the following equalities:
\bea
	\left(\pm \frac{1}{pz^{p-1}}\frac{\pl}{\pl z}\right)\Phi_{p;\bT_q}^\pm = \left(\pm \frac{1}{pz^{p-1}}\frac{\pl}{\pl z}\right)C^\pm(z)\Psi_{p;\bT_q}^\pm =\nonumber\\
	 C^\pm(z) \left(\pm \frac{1}{pz^{p-1}}\frac{\pl}{\pl z}\right)\Psi^\pm_{p;\bT_q} + C^\pm(z)\left( \mp \frac{p-1}{2p z^p} + \sum_{\ell = 1}^q T_{p+\ell}z^{\ell} \right)\Psi^\pm_{p;\bT_q}=\nonumber\\
	 C^\pm(z){\mathcal A}_{p;\bT_q}^\pm \Psi_{p;\bT_q}^\pm \overset{(i)}{=} C^\pm(z)\PR^\pm_{\AR_{p;\bT_q}^\pm}\Psi^\pm_{p;\bT_q} \overset{(ii)}{=} C^\pm(z)\left(\PR^\pm_{\AR_{p;\bT_q}^\pm}\right)_+\Psi^\pm_{p;\bT_q} \overset{(iii)}{=} \nonumber\\
	 C^\pm(z)\left(\left(\sum_{\ell =1}^q T_{p+\ell}\LR_{p;\bT_q}^{\pm \ell/p}\right)_+\Psi_{p;\bT_q}^\pm\right) = \nonumber \\ \left(\left(\sum_{\ell =1}^q T_{p+\ell}\LR_{p;\bT_q}^{\pm \ell/p}\right)_+C^\pm(z)\Psi_{p;\bT_q}^\pm\right) = Q^\pm_{p;\bT_q}\Phi_{p;\bT_q} \label{2.34'}.
\eea
where in $(i)$ and $(iii)$ we used the mappings (\ref{10}), and in $(ii)$ we used (\ref{12}) together with (\ref{generalinvariance}).\\
The first equations in (\ref{LaxNormalized}) and (\ref{mpKP0}) are clearly equivalent. In order to prove the equivalence between the second equations in (\ref{LaxNormalized}) and (\ref{mpKP0}) we observe that the one in (\ref{LaxNormalized}) is of the form
\be
	\pm\frac{\pl}{\pl (z^p)} \Phi^\pm_{p;\bT_q} = \sum_{j = 0}^{q} v_{j}^\pm D^j\Phi^{\pm}_{p;\bT_q},
\label{23}\ee 
where $v_{j}^\pm$ are differential polynomials in the variables $\lbrace \theta_0^\pm,\ldots, \theta_{p-2}^\pm \rbrace$. Then, using the first equation in (\ref{LaxNormalized}), for any $j$ we can write $D^j \Phi^\pm_{p;\bT_q}$ as a linear combination of $\lbrace \Phi^\pm_{p;\bT_q},\ldots,D^{p-1}\Phi^\pm_{p;\bT_q} \rbrace$ with coefficients in the variables\\ $\left\{z^p, D^k \theta_\ell^\pm; \, k \geq 0, \ell = 0,\ldots,p-2\right\}$ so that (\ref{23}) can be rewritten as 
\be
	\pm\frac{\pl}{\pl (z^p)} \Phi^\pm_{p;\bT_q} = \sum_{j = 0}^{p-1} w_{0;j}^\pm D^j\Phi^{\pm}_{p;\bT_q}
\label{23bis}\ee
for some polynomials $w_{0;j}$ in the variables  $\lbrace z^p, D^k \theta_\ell^\pm; \, k \geq 0, \ell = 0;\ldots,p-2\rbrace$. Acting on (\ref{23bis}) with $D^k$ and then using again the first equation in (\ref{LaxNormalized}), we find in the same fashion as above
\be
	\pm\frac{\pl}{\pl (z^p)} D^k\Phi^\pm_{p;\bT_q} = \sum_{j = 0}^{p-1} w_{k;j}^\pm D^j\Phi^{\pm}_{p;\bT_q},\;\quad\,\forall k=1,\ldots,p-1,
\label{23tris}\ee
where again $w_{k;j}$ are polynomials in the variables  $\lbrace z^p, D^k \theta_\ell^\pm; \, k \geq 0, \ell = 0,\ldots,p-2\rbrace$ of at most degree $\left[\frac{q+p-1}{q}\right] = \left[\frac{p-1}q\right]+1$ in $z^p$. Equations (\ref{23bis}),(\ref{23tris}) gives the second equation of (\ref{mpKP0}) with $V_{p;\bT_q}^\pm= (w^\pm_{k,j})_{k,j = 0}^{p-1}$.\\
Finally note that relations (\ref{LaxNormalized}) at $\bt = 0$ amount to
\be
	\LR_{p;\bT_q}^\pm\Psi_{p;\bT_q}^\pm(x,0;z) = z^p\Psi_{p;\bT_q}^\pm(x,0;z)
\label{eigen1}\ee
and
\be
	\AR_{p;\bT_q}^\pm\Psi_{p;\bT_q}^\pm(x,0;z) = P^\pm_{\AR_{p;\bT_q}^\pm}\Psi_{p;\bT_q}^\pm(x,0;z),
\label{eigen2}\ee
this latter coming from (\ref{2.34'}). The two equations (\ref{eigen1}) and (\ref{eigen2}) completely characterize $\Psi^\pm_{p;\bT_q}(x,0;z)$ as an asymptotic series such that $\Psi^\pm_{p;\bT_q}(x,0;z) = 1+\mathcal O(\frac{1}z)$ for $z\rightarrow\infty$, while $\Psi^\pm_{p;\bT_q}(x,0;z)$ characterizes $\WR_{p;\bT_q}^\pm$ by (\ref{11}) and hence $\LR_{p;\bT_q}^\pm$ and $Q_{p;\bT_q}$, yielding the assertion after (\ref{asymPhi}).\qed
\begin{remark}\label{linearityremark}
	We have shown in the proof, using (\ref{9'}), that 
	$$\frac{\partial}{\partial z^p}\hat\Phi^\pm_{p;\bT_q} = \sum_{\ell = 1}^q T_{p+\ell}\frac{\partial}{\partial t_{\ell}}\hat\Phi^\pm_{p;\bT_q}.$$

\end{remark}
Now, given a pseudo--differential operator $A = \sum_i a_iD^i$, recall
$$\mathrm{tr}(A) := D^{-1}a_{-1}.$$
Then, remembering 
$$\LR_{p;\bT_q}^+ = D^p + \sum_{i = 0}^{p-2} \theta_i^+D^i,\quad \Theta^+:=(\theta^+_0,\ldots,\theta_{p-2}^+),$$
we set
$$H_\ell^{(p)} := \frac{p}{p+\ell}\mathrm{tr} \LR_{p;\bT_q}^{+\frac{\ell+p}p},$$
and
$$R_\ell^{(p)} := \left(\frac{\delta H_{\ell}^{(p)}}{\delta\theta_0^+} ,\ldots, \frac{\delta H_{\ell}^{(p)}}{\delta\theta_{p-2}^+} \right),$$
the latter being variational derivatives. Note that by the first symplectic structure, we have \cite{AdlerTrace}
$$J_1R_{\ell}^{(p)} = \left[\left(\LR_{p;\bT_q}^{+\frac{\ell}{p}}\right)_+,\,\LR_{p;\bT_q}^+\right].$$
The well--known Lenard relations \cite{AdlerTrace} are used to recursively compute $R_{\ell}$ through the relation
$$J_1R_{\ell + p}^{(p)} = J_2R_\ell^{(p)},$$
where $J_1$ and $J_2$ are the $(p-1)\times(p-1)$ matrices of the differential operators defining respectively the first and second symplectic structures of the p--Gel'fand--Dickey hierarchy \cite{AdlerTrace}. Now let $\bn$ be the most general solution to the system of differential equations $$J_1\bn = (1,0,\ldots,0)^\T.$$ We can now state:
\begin{proposition}\label{prop2.7}
The string relation in (\ref{SE2}) $$\left[\LR_{p;\bT_q}^+,Q_{p;\bT_q}^+\right] = 1$$
takes the following form of ``Painlev\'e--like'' differential equations for the vector $\Theta^+ = (\theta_0^+,\ldots,\theta_{p-2}^+)$, namely
\be
	\sum_{\ell = 1}^q T_{p+\ell}R_\ell^{(p)} + \bn = 0.
\label{41bis}\ee
\end{proposition}
\proof The string relations take the form
\bea
	0 &=& 1 + \left[Q_{p;\bT_q}^+,\LR_{p;\bT_q}^+\right] = 1+ \sum_{\ell = 1}^q T_{p+\ell}\left[\left(\LR_{p;\bT_q}^{+\frac{\ell}p}\right)_+, \LR_{p;\bT_q}^{+}\right]\\
	   &=& 1+ \sum_{\ell = 1}^q T_{p+\ell} J_1R_\ell^{(p)} = J_1\left(\bn+\sum_{\ell = 1}^q T_{p+\ell}R_\ell^{(p)}\right),
\eea
hence 
$$\sum_{\ell = 1}^q T_{p+\ell}R_\ell^{(p)} + \bn = 0.$$ \qed

\subsection{An example: $p=2$.}

Let's consider the KdV case (p=2). In this case, given any solution of the KdV hierarchy, its Lax operator $\LR^+$ is self--adjoint, hence we are dealing just with one operator that we will denote as
\be
	\LR := \LR^+ = \LR^- = D^2 + 2 y,\quad y:=\pl_x^2\ln\tau
\label{KdVLaxOperator}\ee
We will write the ``Painlev\'e--like'' equations (\ref{41bis}) for the cases $\bT_{2\bar q-1} = (0,0,\ldots,1)$ (all the other ones can be deduced by linearity). The equation (\ref{SE2}) is very conveniently written introducing the universal\footnote{Universal means that they do not depend on the particular solution $\theta$ of the KdV hierarchy we are dealing with. Here $J_1 = D, J_2 = \frac{1}4 D^3 + 2y D + y' $ and $R^{(2)}_{2j+1} = 2\omega_{j+1}$.} Gel'fand--Dickey polynomials $\omega_j(y)$ defined by the equations
\be
	J_1R_{2j+1}^{(2)} = \left[\left(\LR^{j+\frac{1}2}\right)_+,\LR\right] := 2 D\omega_{j+1}
\label{defGDpoly}\ee
and determined recursively by Lenard's recursion $J_1R_{2j+1}^{(2)} = J_2R_{2j-1}^{(2)}:$
\be
	\omega_0 =1,\; D\omega_{j+1} = \left(\frac{1}4 D^3 + 2y D + y'\right)\omega_j.
\label{Lenardrecursion}\ee 
Indeed, using equation (\ref{defGDpoly}), we can rewrite (\ref{SE2}) as
$$2D\omega_{\bar q} = -1$$ so that, integrating once, we get the equations
\be
	2\omega_{\bar q}(y) + x + c_{\bar q} = 0, \quad \bar q\geq 1.
\label{PIhierarchy}\ee
where $c_{\bar q}$ are constants. The first few equations, using (\ref{Lenardrecursion}), read (denoting with a prime the derivative with respect to $x$)
\bea
	\bar q = 1: & 2y + x + c_1 =  0, \label{PI0}\\
	\bar q=2: & 3 y^2 + \dis\frac{1}2 y'' + x + c_2  = 0,\label{PI1}\\
	\bar q=3: & 5 y^3 +\dis\frac{5}2 y y'' + \dis\frac{5}{4} (y')^2 + \dis\frac{1}{8} y^{(iv)} + x + c_3  = 0.\label{PI2}	
\eea
More generally, for $Q_{2;\bT_{2\bar q-1}} = \sum_{j=1}^{\bar q}T_{2j+1}\left(\LR^{+\frac{2j-1}2}\right)_+$, $y(x)$ satisfies
\be
	2\sum_{j=1}^{\bar q} T_{2j+1}\omega_j(y) + x + c = 0.
\label{2.50'}\ee
The equations (\ref{PIhierarchy}) are known as the so--called ``Painlev\'e I hierarchy'' (see for instance \cite{KapaevStokesPI,KapaevSolutionsPI,KudryashovSoukharev} and, for the same approach as the one we used here, \cite{MooreMatrix,MooreGeometrySE} where the same equations are called \emph{massive $(2\bar q-1,2)$ string equations}).\\
We now exhibit $U$ and $V_{2;\bT_{2\bar q -1}}$ (for $\bT_{2\bar q -1}$ given above) as in Theorem \ref{ccSE}:
\be
	U = \left[\begin{array}{cc}
					0 & 1\\
					&\\
					z^2-2y & 0
					\end{array}\right],\quad V_{2;\bT_{2\bar q -1}} = \left[\begin{array}{cc}
															-\frac{1}2 u'_{\bar q-1} & u_{\bar q-1}\\
															&\\
															(z^2 - 2y)u_{\bar q-1}-\frac{1}2u_{\bar q-1}'' & \frac{1}2 u'_{\bar q-1}
															\end{array}\right]
\label{PIMatrices}\ee
where $u_{\bar q} := \sum_{j = 0}^{\bar q} z^{2(\bar q-j)}\omega_j(y)$ and $\omega_j$ are the Gel'fand--Dickey polynomials defined in (\ref{defGDpoly}).\\ The matrices $U$ and $V_{2;\bT_{2\bar q -1}}$ in (\ref{PIMatrices}) give the standard Lax pair for the Painlev\'e I hierarchy (\ref{PIhierarchy}) in the form
\be
	\frac{\pl}{\pl (z^2)} U - DV_{2;\bT_{2\bar q -1}} = \left[V_{2;\bT_{2\bar q -1}} , U\right].
\label{LaxPIa}\ee
The first few matrices $V_{2;\bT_{2\bar q -1}}$ read 
\bea
	V_{2;(1)} = \left[\begin{array}{cc}
					0 & 1\\
					&\\
					z^2-2y & 0
					\end{array}
					\right];\; V_{2;(0,0,1)} = \left[\begin{array}{cc}
					-\frac{y'}2 & z^2 + y\\
					&\\
					z^4-z^2y-2y^2-\frac{1}2 y'' & \frac{y'}2
					\end{array}
					\right];\nonumber\\
					V_{2;(0,0,0,0,1)} = \left[\begin{array}{cc}
					-\left(\frac{z^2}{2}y'+\frac{3}{2} yy' + \frac{1}{8}y'''\right)	&	z^4 + z^2y+\frac{3}{2} y^2 +\frac{1}{4} y''\\		
					f(y) & \left(\frac{z^2}{2}y'+\frac{3}{2} yy' + \frac{1}{8}y'''\right)
					\end{array}
					\right]\label{2.54}\\
					f(y) := z^6 - y z^4 -\left(\frac{y^2}2 + \frac{y''}2 \right) z^2  -3y^3 - 2yy''-\frac{3}2 (y')^2 -\frac{1}8 y'''' =\nonumber\\
					= z^6 - y z^4 -\left(\frac{y^2}2 + \frac{y''}4 \right) z^2 + 2y^3 +\frac{1}2 y y'' - \frac{1}4 (y')^2 + x +c_3\nonumber
\eea
(remark that in the last equality we used (\ref{PI2})). More generally, by linearity, for arbitrary $\bT_{2\bar q - 1} = (T_1,\ldots,T_{2\bar q -1})$ we find
\be
	V_{2;\bT_{2\bar q -1}} = \sum_{j = 1}^{\bar q} T_{2j+1}V_{2;T_{2j-1}},\label{2.53'}
\ee
where $V$ acts on the wave vector function as the derivative with respect to $t_{2j-1}$, namely
$$\frac{\partial}{\partial t_{2j-1}} \hat\Phi^-_{2,\bT_{\bar q}} = V_{2;T_{2j-1}}\hat\Phi^-_{2,\bT_{\bar q}} .$$
\proof
Observe that we have, for $\bT_{2\bar q -1} = (0,\ldots,0,1)$, $Q_{2;\bT_{2\bar q -1}} = (\LR^\frac{2\bar q-1}{2}_{2;\bT_{2\bar q -1}})_+$.
We start proving the following equality (see for instance \cite{BBTbook}), upon setting $\LR = D^2 + \theta$:
\be
	(\LR^\frac{2j+1}{2})_+ = \sum_{i = 0}^j \left(\omega_i D -\frac{1}2 \omega_i' \right)\LR^{j-i}, \; j\geq 0.
\label{Lenard1}\ee
Indeed, using the definition of Gel'fand--Dickey polynomials (\ref{defGDpoly}), we obtain
$$2D\omega_j = [(\LR^{j-\frac{1}2})_+,\LR] = [\LR, (\LR^{j-\frac{1}2})_-] = [D^2 + \theta, (\LR^{j-\frac{1}{2}})_{-1}D^{-1}+(\LR^{j-\frac{1}{2}})_{-2}D^{-2}+\ldots],$$ 
forcing
\be
	(\LR^{j-\frac{1}{2}})_{-} = \omega_j D^{-1} -\frac{1}2 \omega_j' D^{-2}+\ldots .
\label{2.46}\ee
On the other hand, since $\LR = \LR^+$, and using (\ref{2.46}) and $\LR = D^2 +\theta$, conclude:
$$(\LR^{\frac{2j+1}{2}})_+ = (\LR^{\frac{2j-1}{2}})_+\LR + ((\LR^{\frac{2j-1}{2}})_-\LR)_+ = (\LR^{\frac{2j-1}{2}})_+\LR + \omega_j D -\frac{1}2 \omega_j',$$
and this last equation gives (\ref{Lenard1}) by induction on $j \geq 0$. Then we get, using (\ref{Lenard1}), 
\bea
	\frac{\pl \Phi_{2;\bT_{2\bar q+1}}}{\pl (z^2)} = (\LR_{2;\bT_{2\bar q +1}}^{\frac{2\bar q+1}{2}})_+\Phi_{2;\bT_{2\bar q +1}} = \sum_{i = 0}^{\bar q} \left(\omega_i D -\frac{1}{2}\omega_i' \right)\LR^{\bar q-i}_{2;\bT_{2\bar q +1}}\Phi_{2;\bT_{2\bar q +1}} = \nonumber\\
=\sum_{i = 0}^{\bar q} z^{2(\bar q-i)}\left(\omega_i D -\frac{1}{2}\omega_i' \right)\Phi_{2;\bT_{2\bar q +1}} = -\frac{1}2 u'_{\bar q}\Phi_{2;\bT_{2\bar q +1}} + u_{\bar q} D\Phi_{2;\bT_{2\bar q +1}}.
\nonumber\eea
 This latter gives the first line of $V_{2;\bT_{2\bar q -1}}$; the second one is easily obtained by acting on the above equation with $D$ and using the equation\\ $D^2\Phi_{2;\bT_{2\bar q -1}} = (z^2-\theta)\Phi_{2;\bT_{2\bar q -1}}.$ \qed
 
 \section{$(p,q)$-kernels, vertex operators and Virasoro}

We start defining some (integrable) kernels generalizing the one given in \cite{ACvM1}; Definition 3.1. We recall that we denoted with $\Psi_{p;\bT_q}^\pm$ the wave functions associated to $\WR_{p;\bT_q}^\pm$ and with $\Phi_{p;\bT_q}^\pm$ the normalized ones defined by (\ref{normalizedwf}). 
\begin{definition}\label{3.1}
	Given the wave functions $\Psi_{p;\bT_q}^\pm$ we define the integral $(p,q)$-kernels
\bea
	 k^{(p;\bT_q)}_{x,{\mathbf t}}(z,z')&:=&D^{-1}\left(~\Psi^-_{p;\bT_q}(x,{\mathbf t};z)\Psi^+_{p;\bT_q}(x,{\mathbf t};z')\right),\label{defker1}\\
K^{(p;\bT_q)}_{x,{\mathbf t}}(\lb,\lb')&:=& {\rm e}^{-\sum_{\ell = 1}^q\frac{p}{p+\ell}z^{p+\ell}T_{p+\ell}}\left.
 \frac{k^{(p;\bT_q)}_{x,{\mathbf t}}(z,z')}{2\pi p z^{\frac{p-1}{2}}z'^{\frac{p-1}{2}}}
{\rm e}^{\sum_{\ell = 1}^q\frac{p}{p+\ell}z'^{p+\ell}T_{p+\ell}}
\right|_{{z=\lb^{1/p}}\atop{z'=\lb'^{1/p}}}\label{defker2}
\label{defkernels}\eea
\end{definition} 
Note that $K^{(p;(1))}_{x,{\mathbf t}}(\lb,\lb') = K^{(p)}_{x,{\mathbf t}}(\lb,\lb')$ in \cite{ACvM1}. In the following, given a vector $\bT_q=(T_{p+1},\ldots,T_{p+q})$ as in Definition \ref{shifteddefinition}, we extend it to an infinite vector $(T_1,T_2,T_3,\ldots)$ imposing  $T_k =0\;\forall\, k\leq p\;and\; k > p+q$. Since, by (\ref{GrassmannFlow}) and (\ref{shiftedsubspaces}) 
$$\WR^\pm_{p;\bT_q}(\bt) = W^\pm_{p;(1)}(\tilde\bt)\; \mathrm{with}\; \tilde t_j = t_j +\frac{p}{j} T_{j} -\frac{p}{p+1}\delta_{p+1,j}, $$
by (\ref{9'}) we have for all $\bt$ and $\bT_q$:
$$\Psi_{p;\bT_q}^\pm(x,\bt;z) = {\rm e}^{\pm (xz + \sum_{i=1}^\infty t_i z^i)}\frac{\tau_p\left(\bar{\tilde\bt}\mp[z^{-1}]\right)}{\tau_p\left(\bar{\tilde\bt}\right)}$$
where $\tau_p$ denotes the tau function going with $\WR^\pm_p = \WR^\pm_{p;(1)}$ and remember $\bar{\tilde{t_i}} = \tilde{t}_i + \delta_{i1}x$ . Thus, (\ref{defker1}) and (\ref{defker2}) immediately yields the crucial identity

\be
	K^{(p;\bT_q)}_{x,{\mathbf t}}(\lb,\lb') = K^{(p;(1))}_{x,\tilde{\mathbf t}}(\lb,\lb'),\;\mathrm{with}\; \tilde t_j = t_j +\frac{p}{j} T_{j} -\frac{p}{p+1}\delta_{p+1,j}.
\label{identitykernels}\ee
%\proof
%We start observing that we have
%$$\left.K^{(p;\bT_q)}_{x,{\mathbf t}}(\lb,\lb') = -\frac{1}{4\pi^2 i}D^{-1}\left(\Phi^-_{p;\bT_q}(x,\bt;z)\Phi^+_{p;\bT_q}(x,\bt;z')\right)\right|_{{z=\lb^{1/p}}\atop{z'=\lb'^{1/p}}}$$
%so that, in order to prove (\ref{identitykernels}), is enough to prove
%\be
%	\Phi^\pm_{p,\bT_q}(x,\bt;z) =  \Phi^\pm_{p,(1)}(x,\tilde\bt;z),\mathrm{with}\; \tilde t_j = t_j +\frac{p}{j} T_{j} -\frac{p}{p+1}\delta_{p+1,j}.
%\label{identityphi}\ee
%We recall that, by definition, $\WR_{p;\bT_q}^\pm(\bt) = \WR_{p;(1)}^\pm(\tilde\bt).$ 
%The analogue formula for $\Psi_{p;\bT_q}^\pm$ does not hold, because by definition the wave function has asymptotics $$\Psi_{p;\bT_q}^\pm = {\rm e}^{xz}\left(1+\mathcal O(z^{-1})\right);$$
%for $\bt = 0$; rather we have
%$$\Psi_{p;\bT_q}^\pm(x,\bt;z) = {\rm e}^{\mp\sum_{\ell=1}^q\frac{p}{p+\ell}T_{p+\ell}z^{p+\ell}\pm\frac{p}{p+1}z^{p+1}}\Psi_{p;(1)}^\pm(x,\tilde\bt;z)$$
%and this last equation, together with the definition (\ref{normalizedwf}), gives (\ref{identityphi}), hence (\ref{identitykernels}). \qed
For sake of clarity we will report here, slightly rephrasing it, Proposition 3.3 of \cite{ACvM1}. This proposition together with (\ref{identitykernels}) will give us the new Proposition \ref{newprop4} below.
In the following we denote ($:\;:$ means normal ordering)
$$W_i^{(1)}(\bt)=\frac{\pl}{\pl t_i}+(-i)t_{-i},\; W^{(2)}_{\ell}(\bt)=\sum_{i+j=\ell}:W_i^{(1)}(\bt)W_j^{(1)}(\bt):-(\ell+1)W_\ell^{(1)}(\bt)$$ 
and $c_{p,j} = \delta_{1,j}\dis\frac{p^2-1}{12 p^2}$.\\
Now introduce the KP vertex operator
\be
\BX({\mathbf t},y,z):=\frac{1}{z-y}e^{\sum_1^{\iy}(z^i-y^i)t_i}e^{\sum_1^{\iy}(y^{-i}-z^{-i})\frac{1}{i}\frac{\pl}{\pl t_i}}.
\nonumber\ee
Given a p--reduced tau function $\tau(\bt)$ and a disjoint union of intervals $E := \bigcup^r_{i=1}[a_{2i-1},a_{2i}]\subset\BR^+$ define another function $\tau_E({\mathbf t})$:
\be
	\tau_E({\mathbf t}):= {\rm e}^{^{-\dis \mu\displaystyle{\int_{E^{{1}/{p}}}}dz~\BX (t;\om z,\om'z)}}\tau({\mathbf t}).
\label{deftauE}\ee
where $E^{\frac{1}p} := \{x \in \BR^+\;\mathrm{s.t.}\; x^p\in E\}$ and $\omega,\omega'$ are two distinct p--roots of unity.
  
\begin{proposition}\label{prop4}
Consider a disjoint union of intervals $E := \bigcup^r_{i=1}[a_{2i-1},a_{2i}]\subset\BR^+$ and the Fredholm determinant $\det ( \un -2\pi\mu K^{(p;(1))}_{x,{\mathbf t}} \raisebox{1mm}{$\chi$}{}_{E} )$. The following equality is satisfied
 
 \be 
	\det ( \un -2\pi\mu K^{(p;(1))}_{x,{\mathbf t}} \raisebox{1mm}{$\chi$}{}_{E} )= \frac{\tau_{p,E}\left(\bar{\mathbf t}\right)}{\tau_p\left(\bar{\mathbf t}\right)}
\label{44}\ee 
where we recall that  $\bar{t_i} := t_i+x\delta_{1,i}$, while $\tau_{p,E}$ and $\tau_p$ are p--reduced KP tau function satisfying the following Virasoro constraints for every $j\geq 0$:
\be
	\left(\frac{1}{2p}W^{(2)}_{(j-1)p}(\bt')-\frac{p-1}{2p}W_{(j-1)p}^{(1)}(\bt')\right)
	\left\{\begin{array}{l}
		\tau_p({\mathbf t'})\\
		\\
		\tau_{p,E}({\mathbf t'})
	\end{array}\right\} =
	\left\{\begin{array}{l}
		-c_{p,j}\tau_p({\mathbf t'})\\
		\\
		\left(-c_{p,j}+\dis\sum_1^{2r}a^j_i\frac{\pl}{\pl a_i}\right)\tau_{p,E}({\mathbf t'}).
	\end{array}\right\}
\label{Virasoro}\ee
Here $t'_i = t_i + \frac{p}{p+1}\delta_{i,p+1}$.
\end{proposition}
This proposition is exactly the same as Proposition 3.3 of \cite{ACvM1} since, as we said before, $K^{(p;(1))}_{x,{\mathbf t}}(\lb,\lb') = K^{(p)}_{x,{\mathbf t}}(\lb,\lb')$. Also the Virasoro constraints (\ref{Virasoro}) are the same as in (3.9) of \cite{ACvM1} since the shift in the argument of the tau function $\bt\rightarrow\bt'$ makes the missing term $W^{(1)}_{jp+1}$ disappears in (\ref{Virasoro}). Note that the Virasoro constraint holds for \emph{all} $\bt$; hence using the formula (\ref{identitykernels}) we can extend Proposition \ref{prop4} to all the kernels $K^{(p;\bT_q)}_{x,{\mathbf t}}(\lb,\lb')$ simply shifting KP times, which of course shifts $\WR^\pm_p\in\mathrm{Gr}$ by the KP flow. Thus from Proposition \ref{prop4} and formula (\ref{identitykernels}) conclude:
\begin{proposition}\label{newprop4}
Consider a disjoint union of intervals $E := \bigcup^r_{i=1}[a_{2i-1},a_{2i}]\subset\BR^+$ and the Fredholm determinant $\det ( \un -2\pi\mu K^{(p;\bT_q)}_{x,{\mathbf t}} \raisebox{1mm}{$\chi$}{}_{E} )$. The following equality is satisfied
 \be 
	\det ( \un -2\pi\mu K^{(p;\bT_q)}_{x,{\mathbf t}} \raisebox{1mm}{$\chi$}{}_{E} )= \frac{\tau_{p,E}\left(\bar{\mathbf t}^*\right)}{\tau_p\left(\bar{\mathbf t}^*\right)}.
\label{new44}\ee 
 where  $\tau_{p,E}$ and $\tau_p$ are the p--reduced KP tau function described in Proposition \ref{prop4}, satisfying the Virasoro constraints (\ref{Virasoro}) with $\bt'\rightarrow \bt^*$ with
 $t_i^* := t_i+\frac{p}{i}T_i$.
%\be
%	\left(\frac{1}{2p}W^{(2)}_{(j-1)p}(\tilde\bt)-\frac{p-1}{2p}W_{(j-1)p}^{(1)}(\tilde\bt)\right)
%	\left\{\begin{array}{l}
%		\tau({\mathbf t})\\
%		\\
%		\tau_E({\mathbf t})
%	\end{array}\right\} =
%	\left\{\begin{array}{l}
%		-c_{p,j}^{(0)}\tau({\mathbf t})\\
%		\\
%		\left(-c_{p,j}^{(0)}+\dis\sum_1^{2r}a^j_i\frac{\pl}{\pl a_i}\right)\tau_E({\mathbf t})
%	\end{array}\right\}
%\label{newVirasoro}\ee
%with $\tilde t_i = t_i + \frac{p}{i}T_{i}$.
\end{proposition}
Note that Proposition \ref{newprop4} includes as a special case Proposition \ref{prop4}, for $\bT_q = \left(\delta_{p+1,j}\right)_{j=1}^\infty$.
Our kernels $K^{(p;\bT_q)}_{x,{\mathbf t}}$ are given in a rather abstract form in Definition \ref{3.1}; now we want to prove that they can be nicely expressed as integrable kernels \`a la Its--Izergin--Korepin--Slavnov \cite{IIKS}. Given three functions $y(x),z(x);\theta(x)$ we define some differential polynomials $B_i(y,z;\theta)$ by the formulas
\be
	B_0(y,z;\theta) = 0,\quad B_{k+1}(y,z;\theta) = \sum_{\ell = 0}^k D^{k-\ell}y(-D)^\ell (\theta z)
\label{B}\ee
so that the first ones read 
\be
	B_0(y,z;\theta) = 0, \; B_1(y,z;\theta) = \theta y z, \; B_2 = \theta(y'z - yz') - \theta' yz 
\nonumber\ee
and so on. Given a differential operator $L = \sum_{i = 0}^n \theta_i(x)D^i$ we also define the following operation on a pair of functions $y,z$:
	\be
		[y,z]_L := \sum_{i = 0}^n B_i(y,z;\theta_i).
	\ee
The following fact goes back at least to Lagrange (see for instance \cite{Akhiezer-Glazman}); we will denote with $L^*$ the adjoint of $L$.
\begin{lemma}
	Given a differential operator $L := \sum_{i = 0}^n \theta_i(x)D^i$ and two arbitrary differentiable functions $y(x),z(x)$  the following Wronskian equation holds:
	\be
		D [y,z]_L = (Ly)z - y(L^*z).
	\label{lagrangeformula}\ee
\end{lemma}
 \noindent{\it Proof:} \, It suffices, by linearity, to do the case $L_k:=\theta D^{k+1},\, k\geq 0$; which we do by induction on $k$. Compute, using the induction hypothesis,
 \bean
 (\theta D^{k+1}y)z&=&\theta(D^ky')z=y'(-D)^k(\theta z)+DB_k(y',z;\theta)\\
 &=&y(-D)^{k+1}(\theta z)+D(y(-D)^k(\theta z)+B_k(y',z;\theta)),
 \eean
 and so (\ref{lagrangeformula}) holds for $L_k$ if and only if we have
 \be
 B_{k+1}(y,z;\theta)=B_k(y',z;\theta)+y(-D)^k(\theta z)
 \label{6.38}\ee
 with $B_0(y,z;\theta)=0$. On the other hand it is easy to see, again by induction, that formula (\ref{B}) for $B_{k+1}(y,z;\theta)$ is the unique solution of (\ref{6.38}) with $B_0(y,z;\theta)=0$, thus concluding the proof. \qed
 
 \begin{proposition}
 	The integral kernels $K_{x,\bt}^{(p;\bT_q)}(\lb,\lb')$ can be written in the integrable form
	\be
		\left.K_{x,\bt}^{(p;\bT_q)}(\lambda,\lambda') = \frac{\left[\Phi^+_{p;\bT_q}(x,\bt;z'),\Phi^-_{p;\bT_q}(x,\bt;z)\right]_{\LR_{p;\bT_q}^+(x,\bt)}}{ i (z^{'p}-z^{p})}\right|_{{z=\lb^{1/p}}\atop{z'=\lb'^{1/p}}} \!\!\!\!\!\!\!\!= D^{-1}\left(\frac{1}{i}\Phi^-_{p;\bT_q}\Phi^+_{p;\bT_q}\right).
	\label{IIKSkernels}\ee
	where $\Phi^\pm_{p;\bT_q}$ are the normalized eigenfunctions of $\LR^\pm_{p;\bT_q}$ of the Lax system (\ref{LaxNormalized}).
\end{proposition}
\proof
The following equality, coming directly from (\ref{normalizedwf}) and Definition \ref{3.1}, holds:
$$\left.DK_{x;\bt}^{(p;\bT_q)}(\lambda,\lambda') = \frac{1}{i}\left(\Phi_{p;\bT_q}^-(x,\bt;z)\Phi_{p;\bT_q}^+(x,\bt;z')\right)\right|_{{z=\lb^{1/p}}\atop{z'=\lb'^{1/p}}}$$
Multiplying by $(z^{'p}-z^p)$ and using (\ref{LaxNormalized}) and (\ref{lagrangeformula}) we get the following chain of equations:
\bea
	&&(z^{'p}-z^p)DK_{x;\bt}^{(p;\bT_q)}(\lambda,\lambda') =  \nonumber\\
	&&= \frac{1}{i}\left(-\Phi_{p;\bT_q}^+(x,\bt;z')\left(\LR^-_{p;\bT_q}\Phi_{p;\bT_q}^-\right)(x,\bt;z) + \Phi_{p;\bT_q}^-(x,\bt;z)\left(\LR_{p;\bT_q}^+\Phi_{p;\bT_q}^+\right)(x,\bt;z') \right)=\nonumber\\
	&&= \left.\frac{1}{i}D\left[\Phi_{p;\bT_q}^+(x,\bt;z'),\Phi_{p;\bT_q}^-(x,\bt;z)\right]_{\LR^+_{p;\bT_q}(x,\bt)}\right|_{{z=\lb^{1/p}}\atop{z'=\lb'^{1/p}}}
\eea
so that, integrating once, we get (\ref{IIKSkernels}). \qed

\section{Examples from random matrix theory and the Ising model}

\subsection{Higher order Tracy--Widom distributions}

The most interesting case for applications in random matrix theory is for $p = 2$. In this case we have $B_2(y,z;1) = y'z -z'y =: \{y,z\}_x$, $\LR_{2;\bT_q}$ is self--adjoint and so $\Phi^+_{2;\bT_q}(x;z) = \Phi^-_{2;\bT_q}(x;-z)$ as a consequence of the absence of even times in the KdV hierarchy. Hence we get from (\ref{lagrangeformula}) and (\ref{IIKSkernels})
\be
	K_{x,0}^{2;\bT_q}(z'^2,z^2) = \frac{\left\{\Phi^-_{2;\bT_q}(x;-z'),\Phi^-_{2;\bT_q}(x;z)\right\}_x}{i(z'^2-z^2)} = \frac{\hat\Phi_{2;\bT_q}^{-\T}(x;z)\left[\begin{array}{cc} 0 & 1\\-1 & 0 \end{array}\right] \hat\Phi_{2;\bT_q}^-(x;- z')}{i(z^{'2} - z^{2})},
\label{kernelsPI}\ee
where $\hat\Phi_{2;\bT_q}^-$ solves the Lax system for the PI hierarchy
\be
	D\hat\Phi_{2;\bT_q}^- = U \hat\Phi_{2;\bT_q}^-,\quad \frac{\pl }{\pl (z^2)}\hat\Phi_{p;\bT_q}^- = - V_{2,\bT_q}\hat\Phi_{2;\bT_q}^-,
\label{LaxPI}\ee
with $U,V_{2,\bT_q}$ explicitly given in (\ref{PIMatrices}). Hence these are the same kernels studied in \cite{CIK}, with $q = 2\bar q - 1$.\\
In particular consider the case $\bar q = 3$ with $\bT_5 := \left(-\frac{t}{2},0,0,0,\frac{1}{30}\right)$,
\be
	\LR_{2;\bT_5} = \LR_{2;\bT_5}^\pm = D^2 + 2y(x,\bt),\quad \pm Q^\pm_{2;\bT_5} = -\frac{t}2\left(\LR_{2;\bT_5}^{\frac{1}{2}}\right)_+ + \frac{1}{30}\left(\LR_{2;\bT_5}^\frac{5}2\right)_+ .
\ee
From (\ref{PI0})--(\ref{2.50'}) conclude that $y(x,t)$ satisfies the string relation
\be
	0 = x+c_3-t\omega_1(y)+\frac{1}{15}\omega_3(y) = x+c_3-ty+\frac{1}{30}\left(5y^3+\frac{5}2 y y'' + \frac{5}4 y'^2+\frac{1}8y''''\right),\label{4.4}
\ee
 which is often called $PI^2$, being the second member of the Painlev\'e I hierarchy. From (\ref{2.54}) and (\ref{2.53'}) we have
 in this case
\bea
	&V_{2;(-\frac{t}{2},0,0,0\frac{1}{30})} = -\dis\frac{t}2 V_{2;(1)} + \dis\frac{1}{30} V_{2;(0,0,0,0,1)} =\nonumber\\
	\label{4.5}\\ &\dis\frac{1}{240} \left[\begin{array}{cc}
					-\left(4z^2y'+12 yy' + y'''\right)	&	8 z^4 + 8 z^2y+12 y^2 +2 y'' -120t\\	
					&\\	
					f(y) &\left(4z^2y'+12 yy' + y'''\right)
					\end{array}
					\right],\nonumber
\eea
with
$$ f(y) := 8z^6 - 8 y z^4 - \left(4y^2 + 2y'' + 120t \right) z^2 + 16y^3 +4 y y'' - 2 (y')^2 + 240 x + 240 c_3.$$

So, setting $c_3 =0$, this is the same case as the one studied in \cite{ClaeysVanlessen2} (see equations (1.16),(1.17)). Indeed, from (\ref{asymPhi}), conclude that as $z\rightarrow\infty$
\be
	\Phi^+_{2,(-\frac{t}2,0,0,0,\frac{1}{30})}(x,0;-z) = \Phi^-_{2,(-\frac{t}2,0,0,0,\frac{1}{30})}(x,0;z) = \frac{{\rm e}^{-xz+\frac{t}3 z^5 - \frac{1}{105}z^7}}{\sqrt{-4\pi z}}\left(1+\mathcal O(\frac{1}z)\right),
\ee
and so setting 
\be
	\Phi_1(z^2;x,t) := \sqrt{-2\pi} {\rm e}^{-\frac{i\pi}4}\Phi^-_{2;\bT_5}(x,t;z),\quad \Phi_2(z^2;x,t) := \frac{\partial}{\partial x}\Phi_1(z^2,x,t),
\ee
one also finds the $\Phi_j(\xi;s,t)$ appearing in \cite{ClaeysVanlessen2}.\\
More generally, consider the general cases from \cite{CIK}, $q = 4k+1$:
\be
	\LR_{2;\bT_q} := \LR^\pm_{2;\bT_q} = D^2 + 2y(x,\bT_q), \quad \pm Q^\pm_{2;\bT_q} = \sum_{j = 1}^{2k+1} T_{2j+1}\left(\LR_{2;\bT_q}^{j-\frac{1}2}\right)_+,
\label{4.8}\ee
Again the string relation (\ref{2.50'}) is a member of the $PI$ hierarchy, namely $PI^{2k}$. Not only do we have the Lax system for the $PI$ hierarchy (\ref{LaxPI}), with $V_{2;\bT_{4k+1}}$ given by (\ref{2.53'}), but in addition (setting $t_j = \frac{2}j T_j$) we also find from Remark \ref{linearityremark} and (\ref{2.53'}) that 
\be
	\frac{\partial \hat\Phi^+_{2;\bT_{4k+1}}(x,\bt;z)}{\partial t_{2j+1}} = V_{2;\bT_{2j+1}} \hat\Phi^+_{2;\bT_{4k+1}}(x,\bt;z)
\ee
with $V_{2;\bT_{2j+1}}$ given by (\ref{PIMatrices}). Now as $z\rightarrow\infty$

\be
	\Phi^+_{2;\bT_{4k+1}}(x;-z) = \frac{1}{\sqrt{- 4\pi z}}{\rm e}^{-xz-2\sum_{j=1}^{2k+1}\frac{T_{2j+1}}{2j+1}z^{2j+1}}\left(1+\mathcal O\left(\frac{1}z\right)\right)
\ee
and as before set
\be
	\Phi_1(z^2,d_1T_3,\ldots,d_{2k+1}T_{4k+3}) = \sqrt{-2\pi} {\rm e}^{-i\frac{\pi}4}\Phi^+_{2;\bT_{4k+1}}(x,0;-z),\quad \Phi_2 = \frac{\partial}{\partial x}\Phi_1
\ee
for appropriate constants $d_1,\ldots,d_{2k+1}$ to find the $\Phi_j(\xi;s,t)$ of \cite{CIK}.

\subsection{The Ising model: $p = 3$ and $4$}

Next we consider the case, coming from a realization as a two--matrix model of the critical Ising model (see \cite{DFGZJ} and references therein). The model comes from the string equation
$$[\LR^+_{3;\bT_4},Q^+_{3;\bT_4}] = 1$$
for the Lax operators
\bea
	\LR_{3;\bT_4}^+ &=& (D^2 - u)^{\frac{3}{2}} + \frac{3}{2} w = D^3 - \frac{3}2 uD + \frac{3}4 (2w-u'),\nonumber\\
	Q^+_{3;\bT_4} &=&\!\!\!\! (\LR_{3;\bT_4}^{+\frac{4}3})_+ + T_5(\LR_{3;\bT_4}^{+\frac{2}3})_+\label{LaxIsing},\\
				&=& D^4 - (D^2u + uD^2) + (wD+Dw)+\frac{1}{2}u^2-\frac{1}6 u^{(iv)}+T_5(D^2-u)\nonumber
\eea
with
$$u=-\frac{1}2\pl_x^2\log\tau_3, \; w = \pl_x\pl_2\log\tau_3,$$
so that, in our notation, we have $\bT_4 = (0,T_5,0,1)$. The string equation (\ref{SE2}), using Proposition \ref{prop2.7}, has a solution of the form 
\be\left\{\begin{aligned}
	&\dis\frac{1}2 w''-\dis\frac{3}2 uw + \dis\frac{3}2 T_5 w + t_2 = 0\\
	\\
	&\dis\frac{1}{12} u^{(iv)}-\dis\frac{3}4 uu''-\frac{3}{16} (u')^2 + \dis\frac{1}4 u^3 -\dis\frac{1}{4} T_5(3u^2 - u'') + \dis\frac{3}2 w^2 +x = 0.\end{aligned}\right.
\label{string43}\ee
The last case we consider is related to the so--called tricritical Ising  model (see again \cite{DFGZJ}) and comes from the string equation 
$$[\LR^+_{4;\bT_5},Q^+_{4;\bT_5}] = 1$$
for the Lax operators
\bea
	\LR_{4;\bT_5}^+& = &(D^2 - u)^2 + wD + Dw +v,\nonumber\\
	\label{LaxtriIsing}\\
	Q^+_{4;\bT_5}& =&(\LR_{4;\bT_5}^{+\frac{5}4})_+ = ((D^2-u)^\frac{5}{2})_++\frac{5}4(wD^2 + D^2w)+\frac{5}8(vD+Dv)-\frac{5}4 uw,\nonumber
\eea
with
\be u=-\frac{1}2\pl_x^2\log\tau, \; w = \pl_x\pl_2\log\tau,\; v = \frac{4}3\pl_x\pl_3\ln\tau-\frac{1}3\pl_x^4\ln\tau+2(\pl_x^2\ln\tau)^2\label{uvw},\ee
so that, in our notation, we have $\bT_5 = (0,0,0,0,1)$. The string equation (\ref{SE2}) leads to the system of ODEs (recall that we denoted the $j$-th Gel'fand--Dickey polynomial $\omega_j$, see (\ref{defGDpoly}))
\be\left\{\begin{aligned}
	&2\omega_3\left(-\dis\frac{u}{2}\right) + \dis\frac{5}8 v''- \dis\frac{5}4uv+\dis\frac{5}4w^2+\dis\frac{3}2 t_3 = 0 \\
	\\
	&\dis\frac{1}2w^{(iv)} - \dis\frac{5}4\left(uw\right)''- \dis\frac{5}4 u w''-\dis\frac{5}2vw+\frac{5}4 u^2w - 4t_2 = 0\\
	\\
	&4\omega_4\left(-\dis\frac{u}2\right)+\dis\frac{1}{16}v^{(iv)}+\dis\frac{5}{8}v^2+\dis\frac{15}8u^2v-\dis\frac{5}8(uv''+u'v'+vu'')-\frac{5}4 ww''+\frac{5}4 w^2 u -\frac{3}2 t_3 u + t_1 = 0.&&\\
	\end{aligned}\right.
\label{stringtriIsing}\ee

\section{Nonlinear PDEs for $(p,q)$--kernels.}

As it was the case for the previous article \cite{ACvM1}, also in this case Proposition \ref{newprop4} gives a method, combining it with the equations of the KP hierarchy,  to derive nonlinear PDEs for the Fredholm determinants $\left.\det ( \un -2\pi\mu K^{(p;\bT_q)}_{x,{\mathbf t}} \raisebox{1mm}{$\chi$}{}_{E} )\right | _{\mathrm{most}\,t_i = 0}$, which however explicitly depend on the solutions to the string relations. In this section we shall work on a few examples of physical interest. We start recalling a few facts, see  \cite{ACvM1}. In the following, for every $i\in\N$, we denote $\pl_i := \frac{\pl}{\pl t_i}$.

\begin{lemma}\label{lemma6}The bilinear identity for KP (\ref{BHE}) generates two strings of Hirota relations,\footnote{We recall the standard notation for the Hirota symbol of two functions $f$ and $g$, associated with any polynomial of many variables
$$p(\pl_1,\pl_2,\ldots)f \circ g := p\left(\frac{\pl}{\pl t_1},\frac{\pl}{\pl t_2},\ldots\right)f(t_1+y_1,t_2+y_2,\ldots)g(t_1-y_1,t_2-y_2,\ldots)\big\vert_{\{y_i\}=0}.$$ Also the $\bt'$ here has nothing to do with the one in Proposition \ref{prop4}.}
\be
\begin{array}{lll} 
0&=&\dis\oint_{\iy}\dis\frac{dz}{2\pi i}\tau({\mathbf t}-[z^{-1}]\tau({\mathbf t'}+[z^{-1}])e^{\sum_1^{\iy}z^i(t_i-t'_i)}\big\vert_{t\mapsto t+\frac{1}{2}y\atop{t'\mapsto t-\frac{1}{2}y}}\\
\\
&=&\sum_{j=0}^{\iy}p_j(y)p_{j+1}(\pl_{\mathbf t})e^{-\frac{1}{2}\sum^{\iy}_1y_{\ell}\pl_{\ell}}\tau\circ\tau\\
\\
&=&\sum_{\ell =1}^{\iy}y_{\ell}\left(p_{\ell +1}(\pl_{\mathbf t})-\frac{1}{2}{\pl_1\pl_{\ell}}\right)\tau\circ\tau\\
\\
& &+\sum_{\ell =2}^{\iy}y_1y_{\ell -1}\left(p_{\ell +1}(\pl_{\mathbf t})-\frac{1}{4}
\pl_2\pl_{\ell -1}-\frac{1}{2}\pl_1p_{\ell}(\pl_{\mathbf t})\right)   \tau\circ\tau +\mathbf{O}(y_i^3),
\end{array}
\label{67}\ee
which are independent, for $\ell\geq 5$. The first string, denoted symbolically by $\BY_{\ell}$, is the standard KP hierarchy and we will denote twice the second one minus twice the first one by $\BY_{1,\ell -1}$
\be
\BY_{\ell}:\Bigl(p_{\ell +1}(\pl_{\mathbf t})-\frac{1}{2}\pl_1\pl_{\ell}\Bigr)\tau\circ\tau=0,  ~~~\BY_{1,\ell -1}:\Bigl(\pl_1\pl_{\ell}-\frac{1}{2}\pl_2\pl_{\ell -1}-\pl_1p_{\ell}(\pl_{\mathbf t})\Bigr)\tau\circ\tau=0.
\label{68}\ee

\end{lemma}

 \begin{lemma}\label{A1} The Hirota symbols corresponding to the coefficients of Lemma \ref{lemma6}, with the noncontributing odd terms removed are, up to a constant, as follows
\be
\begin{array}{lll} 
%\BY_3& :&-4\pl_1\pl_3+3\pl_2^2+\pl_1^4\\
%\\
\BY_4&: &-3\pl_1\pl_4+2\pl_2\pl_3+\pl_2\pl_1^3\\
\\
%\BY_5&: &\frac{1}{4}\pl_2\pl_4-\frac{3}{5}\pl_1\pl_5+\frac{1}{9}\pl^2_3+\frac{1}{9}\pl_1^3\pl_3+\frac{1}{8}\pl_1^2\pl^2_2+\frac{1}{360}\pl_1^6\\
%\\
\BY_{1,4}& :&-\frac{1}{8}\pl_2\pl_4+\frac{1}{10}\pl_1\pl_5+\frac{1}{18}\pl_3^2-\frac{1}{36}\pl_1^3\pl_3-\frac{1}{360}\pl^6_1\\
\\
4\BY_{1,4}+10\BY_5& :&\frac{1}{2}\pl_2\pl_4-2\pl_1\pl_5+\frac{2}{3}\pl_3^2+\frac{1}{3}\pl^3_1\pl_3+\frac{1}{2}\pl_1^2\pl^2_2,
\end{array}
\label{69}\ee
whose action on $\tau\circ\tau$ yields the following differential equations for $U=\log\tau$:

\be
\begin{array}{lll} 
%\BY_3& :&\pl^4_1U+6(\pl_1^2U)^2+3\pl^2_2U-4\pl_1\pl_3U=0\\
%\\
\BY_4&: &-3\pl_1\pl_4U+2\pl_2\pl_3U+\pl_1^3\pl_2U+6(\pl_1^2U)(\pl_1\pl_2U)=0\\
\\
%\BY_5&: &-\dis\frac{108}{5}\pl_1\pl_5U+\dis\frac{1}{10}\pl_1^6U+6(\pl^2_1U)^3+3(\pl_1^4U)(\pl^2_1U)\\
%\\
%& &+~9\pl_2\pl_4U+4\pl^2_3U+4\pl_1^3\pl_3U+24(\pl_1^2U)(\pl_1\pl_3 U)\\
%\\
%& &+~9(\pl_1^2U)(\pl_2^2U)+\frac{9}{2}\pl_1^2\pl_2^2U+18(\pl_1\pl_2U)^2\\
%\\
\BY_{1,4}& :&-\dis\frac{36}{5}\pl_1\pl_5U+\dis\frac{1}{5}\pl^6_1U+12(\pl_1^2U)^3+6(\pl^4_1U)(\pl^2_1U)+9\pl_2\pl_4U\\
\\
& &-~4\pl^2_3U+2\pl^3_1\pl_3U+12(\pl^2_1U)(\pl_1\pl_3U)=0\\
\\
4\BY_{1,4}+10\BY_5& :&-4\pl_1\pl_5U+\pl_2\pl_4U+\frac{4}{3}\pl_3^2U+\dis\frac{2}{3}\pl_1^3\pl_3U+4(\pl_1^2U)(\pl_1\pl_3U)\\
\\
& &+~\pl^2_1\pl_2^2U+4(\pl_1\pl_2U)^2+2(\pl_1^2U)(\pl_2^2U)=0.
\end{array}
\label{70}\ee
\end{lemma}
We can start with our examples. In the following, given an ordered collection of points $a_1,a_2,\ldots,a_m$, we denote with $E$ the collection of intervals with $a_i$ as endpoints; namely if $m=2k$ is even $E := \cup_{i=1}^k [a_{2i-1},a_{2i}]$; otherwise, if $m=2k+1$ is odd, $E  := \left(\cup_{i = 1}^{k} [a_{2i-1},a_{2i}]\right) \cup [a_{2k+1},\infty)$ or $E  := (-\infty,a_1]\cup \left(\cup_{i = 1}^{k} [a_{2i},a_{2i+1}]\right)$. Also we will denote $\pl := \sum_{i = 1}^m \frac{\pl}{\pl a_i}$ and $\vr:= \sum_{i = 1}^m a_i\frac{\pl}{\pl a_i}$. When a function $U$ depends on the endpoints $a_i$ of $E$ we write $U=U(E)$.
\begin{proposition}\label{ClaeysVanlessen}
Let us consider the $(2,3)$-kernel (\ref{kernelsPI}) with $\bT_5 = (-\frac{t}2,0,0,0,\frac{1}{30})$ as in \cite{ClaeysVanlessen2}. Then we have that
	$U(E,x,t) := \log\det ( \un -2\pi\mu K^{(2;\bT_5)}_{x,{0}} \raisebox{1mm}{$\chi$}{}_{E} )$ satisfies the PDE

\be
\left\lbrace 60 \pl\pl_xU\!\! +\! 30t \pl_x^2U\!\! -\!\! 6\pl_t^2U\!\! +\! \pl_t\pl_x^2U\!\! +\! 6\pl_x^2U\pl_x\pl_tU\!\! +\! 6y\pl_x\pl_tU,\pl_x^2U\right\rbrace_x\!\!+6\!\left(\pl_x^2U\right)^2\!\pl_ty =0;
\label{6.64bis}\ee
with $y=y(x,t)$ the solution to the string relation (\ref{4.4}), namely\\
$$\frac{1}6 y^3 + \frac{1}{24}(\pl_x y)^2 + \frac{1}{12}y\pl_x^2y + \frac{1}{240}\pl_x^4y-ty + x + c= 0.$$
For the case of \cite{ClaeysVanlessen2}, $E=[s,\infty), \pl = \pl_s,\mu = \frac{i}{2\pi}, c=0$ and $$U \rightarrow 0 \quad \mathrm{as}\quad s\rightarrow\infty,\quad\quad U \sim \left(\frac{5}{16}\right)^2\frac{s^7}{7}, \quad s\rightarrow - \infty.$$
\end{proposition}
 \noindent{\it Proof:} \, By (\ref{Virasoro}) and (\ref{new44}) $\tau_E$ and $\tau$ satisfy (we drop the $p=2$ from $\tau_p$)
 
 \be
 	 \frac{1}{4}W_{-2}^{(2)}(\bt^*)\tau_E(\bt^*) = \pl\tau_E(\bt^*),\quad \frac{1}{4}W_{-2}^{(2)}(\bt^*)\tau(\bt^*) = 0
\nonumber\ee
with  
\be
	t_1^*=x, t_3^*=t_3-\frac{t}3,  t_5^*=t_5, t_7^*=t_7+\frac{1}{105}, \nonumber
\ee
 so that
 \bean
 \pl \tau_E &=&  \frac{1}{2} \Big(3t_3\pl_x+5t_5\pl_3+7t_7\pl_5+9t_2\pl_7+\ldots\Big)\tau_E + \frac{1}{30} \pl_5\tau_E -\frac{t}2\pl_x\tau_E + \frac{x^2}{4}\tau_E,\\
 0 &=&  \frac{1}{2} \Big(3t_3\pl_x+5t_5\pl_3+7t_7\pl_5+9t_2\pl_7+\ldots\Big)\tau + \frac{1}{30} \pl_5\tau-\frac{t}2\pl_x\tau + \frac{x^2}{4}\tau,
  \eean
  hence $g:=\ln\tau_E$ and $g_0:=\ln\tau$ satisfy
  \bean
  \left(\pl+\frac{t}{2}\pl_x\right)g &=& \frac{1}{30}\pl_5 g+\frac{1}{2}\Big(3t_3\pl_x+5t_5\pl_3+\ldots \Big)g+\frac{x^2}{4},\\
  \frac{t}{2}\pl_xg_0 &=& \frac{1}{30}\pl_5 g_0+\frac{1}{2}\Big(3t_3\pl_x+5t_5\pl_3+\ldots \Big)g_0+\frac{x^2}{4}.
 \eean
 Thus
  \bean
\pl_x\left(\pl+\frac{t}{2}\pl_1\right)g&=&\frac{1}{30}\pl_x\pl_5 g+\frac{1}{2}\left(3t_3\pl_x+5t_5\pl_3+\ldots \right)\pl_xg+\frac{x}{2},\\
  \frac{t}{2}\pl_x^2 g_0&=&\frac{1}{30}\pl_x\pl_5 g_0+\frac{1}{2}\left(3t_3\pl_x+5t_5\pl_3+\ldots \right)\pl_xg_0+\frac{x}{2},
  \eean
  and so on the locus $\LR :=\{t_3=t_5=t_7=t_9= \ldots =0\}$,
  
  \bea
    \pl_1\pl_5 g&=&30\pl_1\left(\pl+\frac{t}{2}\pl_1\right)g-15x,\label{6.65a}\\
    \pl_1\pl_5 g_0&=&+15 t\pl_1g_0-15x,
  \label{6.65b}
  \eea
  while by Lemma (\ref{A1}) applied to the $p=2$ case we have that $g$ and $g_0$ both satisfy (denoting with $h$ one of the two):
  
  \be
  4\BY_1\BY_4+10\BY_5:-4\pl_1\pl_5h+\frac{4}{3}\pl^2_3h+\frac{2}{3}\pl^2_1\pl_3h+4(\pl^2_1h)(\pl_1\pl_3h)=0.
  \label{6.66}\ee
  Substituting (\ref{6.65a}), (\ref{6.65b}) in (\ref{6.66}) we find
  \bea
  -120\pl_1\pl g-60t\pl^2_1g+\frac{4}{3}\pl^2_3g+\frac{2}{3}\pl^2_1\pl_3g+4(\pl^2_1g)(\pl_1\pl_3g) + 60 x&=&0, \label{6.67a}\\
  -60t\pl^2_1g_0+\frac{4}{3}\pl^2_3g_0+\frac{2}{3}\pl^2_1\pl_3g_0+4(\pl^2_1g_0)(\pl_1\pl_3g_0) + 60x &=&0,
  \eea
  so that taking the difference of the two equations, we find that $U=g-g_0$ satisfies the equation (setting $\pl_3=-3\pl_t$)
 \be
 	-120\pl_x\pl U-60t\pl^2_1U+12\pl^2_t U - 2\pl^2_x\pl_t U - 12 \pl_x^2 U\pl_x\pl_t U - 12 \pl_x\pl_t g_0\pl_x^2 U - 12 \pl_x^2 g_0\pl_x\pl_t U =0
 \ee
 and this equation (or rather the same one divided by by 2) yields (\ref{6.64bis}) upon dividing by $-6\pl_x^2 U$ and differentiating with regard to $x$, since $\pl_x^2g_0 = y$ (see (\ref{KdVLaxOperator})). \qed
 We could also derive PDEs for the cases (4.8) of \cite{CIK} in a similar fashion.
\begin{proposition}\label{criticalIsing}
Let us consider the $(3,4)$-kernel (\ref{IIKSkernels}) with $\bT_4 = (0,T_5,0,1)$, going with the critical Ising model. Then we have that
	$$\left.V(E,x,t_2) := \log\det ( \un -2\pi\mu K^{(3;\bT_4)}_{x,{\bt}} \raisebox{1mm}{$\chi$}{}_{E} )\right | _{\underset{i\neq 2}{t_i = 0}}$$ satisfies the PDE
\be
3T_5\pl_2\pl_x V -3\pl\pl_xV+\pl^3_x\pl_2V+6(\pl^2_xV)(\pl_x\pl_2V)-12u(\pl_x\pl_2V)+6w\pl^2_xV=0,
\label{6.64}\ee
with $u,w$  satisfying the string relations  (\ref{string43}), and $V\longrightarrow 0$ as $E\longrightarrow\emptyset$.
\end{proposition}
\proof
By (\ref{Virasoro}) and (\ref{new44}) $\tau_E$ and $\tau$ satisfy (again we drop the $p=3$ from $\tau_p$)
  
\bea
  \left(\frac{1}{6}W_{-3}^{(2)}(\bt^*)-c_{3}\right)\tau_E(\bt^*) = \pl\tau_E(\bt^*),\quad \left(\frac{1}{6}W_{-3}^{(2)}(\bt^*)-c_{3}\right)\tau(\bt^*) = 0
\eea
  with
  $$t_1^* = x, t_2^* = t_2, t_4^* = t_4, t_5^* = t_5 +\frac{3}5 T_5, t_7^* = t_7 + \frac{3}7, t_i^* = t_i\quad\forall i\geq 8.$$
 so that, setting $g=\ln\tau_E$ and $g_0 = \ln\tau$ we get equations
  \bea
  \pl g&=&\frac{1}{3}(4t_4\pl_x+5t_5\pl_2+7t_7\pl_4+\ldots)+\pl_4g + T_5\pl_2g,\nonumber
  \nonumber\\
  0&=& \frac{1}{3}(4t_4\pl_x+5t_5\pl_2+7t_7\pl_4+\ldots)+\pl_4g_0 + T_5\pl_2g_0,\nonumber
  \eea
  and so on the locus $\LR :=\{t_4=t_5=t_7= \ldots =0\}$ we get
  \be
  \pl_x\pl_4g = \pl_x\left(\pl-T_5\pl_2\right)g, \quad   \pl_x\pl_4g_0 = \pl_x\left(\pl-T_5\pl_2\right)g_0.
  \label{6.71}\ee
  While by Lemma \ref{A1}, using the equation $\BY_4$ combined with (\ref{6.71}), we obtain that $g$ and $g_0$ satisfies
  \bea
  	 \pl^3_x\pl_2g+6(\pl^2_xg)(\pl_x\pl_2g) &=& 3\pl_x\left(\pl-T_5\pl_2\right)g,\nonumber\\
	 \nonumber\\
	  \pl^3_x\pl_2g_0+6(\pl^2_xg_0)(\pl_x\pl_2g_0) &=&  3\pl_x\left(\pl-T_5\pl_2\right)g_0,\nonumber
  \eea
  and from the difference of these two equations we obtain (\ref{6.64}), using $V=g-g_0$ and the definitions of $u,w$ in (\ref{LaxIsing}).
\qed
\begin{proposition}\label{proptriIsing}
Let us consider the $(4,5)$-kernel (\ref{IIKSkernels}) with $\bT_5= (0,0,0,0,1)$ going with the tricritical Ising model. Then we have that
	$$\left.W(E,x,t_2,t_3) := \log\det ( \un -2\pi\mu K^{(4;\bT_5)}_{x,{\bt}} \raisebox{1mm}{$\chi$}{}_{E} )\right | _{\underset{i\neq 2,3}{t_i = 0}}$$ satisfies the PDE
\bea
 \frac{1}5 \pl_x^6W-4\pl_3^2W+2\pl_x^2\pl_3W+12\left(\pl_x^2W\right)^3+6\left(\pl_x^4W\right)\left(\pl_x^2W\right)+12\left(\pl_x^2W\right)\left(\pl_x\pl_3W\right)\nonumber \\
-72 u \left(\pl_x^2W\right)^2-12u(\pl_x^4W)-24u(\pl_x\pl_3W)+\left(9v+72u^2-18u''\right)\pl_x^2W=\frac{36}5\pl\pl_x W, \nonumber\\
\label{PDEtriIsing}\eea
with $u,v$  satisfying the string relations  (\ref{stringtriIsing}) (together with $w$).
\end{proposition}
\proof
By (\ref{Virasoro}) and (\ref{new44}) $\tau_E$ and $\tau$ satisfy (again we drop the $p=4$ from $\tau_p$)
  
\bea
  \left(\frac{1}{8}W_{-4}^{(2)}(\bt^*)-c_{4}\right)\tau_E(\bt^*) = \pl\tau_E(\bt^*),\quad \left(\frac{1}{8}W_{-4}^{(2)}(\bt^*)-c_{4}\right)\tau(\bt^*) = 0
\eea
  with $t_1^* = x, t_5^* = t_5+\frac{4}5 $ and $t^*_i = t_i, i\neq 1,5$ so that, setting $g=\ln\tau_E$ and $g_0 = \ln\tau$ we get equations
  \bea
  \pl g&=&\frac{1}{4}(5t_5\pl_x+6t_6\pl_2+7t_7\pl_3+\ldots)+\pl_5g,\nonumber
  \nonumber\\
  0&=& \frac{1}{4}(5t_5\pl_x+6t_6\pl_6+7t_7\pl_3+\ldots)+\pl_5g_0,\nonumber
  \eea
  and so on the locus $\LR :=\{t_5=t_6=t_7= \ldots =0\}$ we get
  \be
  \pl_x\pl_5g = \pl_x\pl g, \quad   \pl_x\pl_5g_0 = 0 .
  \label{6.71bis}\ee
  While by Lemma \ref{A1} using the equation $\BY_{1,4}$ combined with (\ref{6.71bis}), we obtain that $g$ and $g_0$ satisfies
  \be\begin{aligned}
  	 &\frac{1}5\pl_x^6g+12\left(\pl_x^2 g\right)^3 + 6 (\pl_x^4g)(\pl_x^2 g) - 4\pl_3^2g+2\pl_x^2\pl_3 g+ 12(\pl_x^2 g)(\pl_x\pl_3g) = \frac{36}5\pl_x\pl g, \nonumber\\
	 &\nonumber\\
	  &\frac{1}5\pl_x^6g_0+12\left(\pl_x^2 g_0\right)^3 + 6 (\pl_x^4g_0)(\pl_x^2 g_0) - 4\pl_3^2g_0+2\pl_x^2\pl_3 g_0+ 12(\pl_x^2 g_0)(\pl_x\pl_3g_0) =  0, \nonumber
  \end{aligned}\ee
  and taking the difference of these two equations we obtain (\ref{PDEtriIsing}), using $W=g-g_0$ and the definitions of $u,w,v$ in (\ref{uvw}).
\qed

\bibliographystyle{plain}
\bibliography{/Users/mattiacafasso/Documents/BibDeskLibrary.bib}

\end{document}